\documentclass[a4paper,11pt]{article}
\pdfoutput=1 

\usepackage{gensymb, xcolor, rotating}
\usepackage{amsmath}
\usepackage{bm}
\usepackage{jheppub} 

\usepackage[T1]{fontenc} 
\usepackage{hyperref}

\definecolor{nicered}{rgb}{0.7,0.1,0.1}

\title{
Indirect Detection of Dark Matter Annihilating into Dark Glueballs
}

\author{David Curtin}
\author{and Caleb Gemmell}
\affiliation{Department of Physics, University of Toronto, Toronto, ON M5S 1A7, Canada}

\emailAdd{dcurtin@physics.utoronto.ca}
\emailAdd{caleb.gemmell@mail.utoronto.ca}

\abstract{
We examine indirect detection of dark matter that annihilates into dark glueballs, which in turn decay into the Standard Model via a range of portals.
This arises if the dark matter candidate couples to a confining gauge force without light flavours, representative of many possible complex dark sectors. Such Hidden Valley scenarios are being increasingly considered due to non-detection of minimal models as well as theoretical motivations such as the Twin Higgs solution to the little hierarchy problem. 
Study of dark glueballs in indirect detection has previously been hampered by the difficulty of modeling their production in dark showers. We use the recent \texttt{GlueShower} code to produce the first constraints on dark matter annihilating via dark glueballs into the Standard Model across photon, antiproton, and positron channels. We also fit the Galactic Centre Excess and use this observation, combined with other astrophysical constraints, to show how multi-channel observations can constrain UV and IR details of the theory, namely the exact decay portal and hadronization behaviour respectively. This provides unique complementary discovery and diagnostic potential to Hidden Valley searches at colliders.
It is interesting to note that thermal WIMPs annihilating to $\mathcal{O}(10~\mathrm{GeV})$ dark glueballs and then the SM via the Twin-Higgs-like decay portal can account for the GCE while respecting other constraints. }

\begin{document} 
\maketitle
\flushbottom

\section{Introduction}
\label{sec:intro}

The nature of dark matter (DM) is one of the biggest ongoing mysteries in the current era of particle physics. While decades of experiments have begun to whittle away at the parameter space of traditional WIMP-like DM, a lack of any signal of new physics across collider, direct detection and indirect detection searches has motivated interest in studying a wider variety of possible models. Inspired by the complexity we observe in our own Standard Model (SM) and the large abundance of DM relative to the SM, significant effort has recently focused on the possibility of \emph{complex dark sectors} in which dark matter and other states may interact via new dark forces.

In general, the term `dark sectors' refers to the case when we consider additional fields that can couple significantly to each other, but are uncharged under the SM. One such framework of dark sector theories are Hidden Valley (HV) models \cite{Strassler:2006im}. Generically, the most common example includes SM-singlet fields charged under a new dark $SU(N_c)$, see e.g. \cite{Kang:2008ea,Bai:2013xga,Renner:2018fhh,Mies:2020mzw}. Specific models such as Mirror Twin Higgs \cite{PhysRevLett.96.231802}, Fraternal Twin Higgs \cite{Craig:2015pha}, Folded SUSY \cite{Burdman:2006tz}, and many more all fall into this bracket \cite{Barbieri:2005ri,Chacko:2005vw,Cai:2008au,Poland:2008ev,Cohen:2018mgv,Cheng:2018gvu}. Interactions with the SM are possible through portal couplings \cite{Holdom:1985ag,Patt:2006fw,Falkowski:2009yz} that are very weak or highly suppressed at low energies. 
A class of dark sectors called Neutral Naturalness models, which are related to the SM via a discrete symmetry and include the Mirror Twin Higgs, can solve the Little Hierarchy Problem while evading conventional LHC searches. This makes the study of dark sectors with qualitative similarities to the complexity of the SM particularly motivated.

When dark quarks or gluons in these confining dark sectors are produced in high energy events, they will have to shower and eventually hadronize, analogous to SM QCD. This will lead to high multiplicity events equivalent to jets we observe at colliders. We refer to these events as `dark showers' \cite{Cohen:2020afv,Knapen:2021eip,Albouy:2022cin}. A range of possible phenomena that arise from dark showers have been studied in the context of colliders, such as semi-visible \cite{Cohen:2015toa,Cohen:2017pzm,CMS:2021dzg} and emerging jets \cite{Schwaller:2015gea,Linthorne:2021oiz,CMS:2018bvr}. Dark showers have also been studied in the context of indirect detection, such as how they explain the Galactic Centre Excess (GCE) \cite{Freytsis:2014sua,Freytsis:2016dgf}. The effect of how phenomenologically similar cascade decays change constraints for DM searches has also been studied \cite{Elor:2015tva,Elor:2015bho}. While not a confining model, the phenomena of high multiplicity softened events is related.

In the $N_f=0$ limit, these dark showers will only contain a variety of dark glueball states \cite{Morningstar:1999rf,Teper:1998kw,Lucini:2010nv,Athenodorou:2021qvs,Yamanaka:2021xqh}. This is an important region of HV parameter space as it commonly appears in neutral naturalness models, but has so far been almost entirely unstudied due to the inapplicability of SM-like hadronization models~\cite{Amati:1979fg,Andersson:1983ia,Webber:1983if} in the $N_f=0$ limit. In this work we  extend the study of DM annihilating to dark showers to include this possibility, making use of recent advances in simulating pure-glue hadronization via the \texttt{GlueShower} code~\cite{Curtin:2022tou}.

Previously dark glueballs have been considered as dark matter candidates themselves \cite{Faraggi:2000pv,Boddy:2014yra,Boddy:2014qxa,GarciaGarcia:2015fol,Soni:2016gzf,Soni:2017nlm,Forestell:2016qhc,Forestell:2017wov,Jo:2020ggs}, but here we only focus on the case that the DM couples to the dark glueball sector, and how this affects the observed annihilation spectra compared to the traditional WIMP expectation. Additionally, we are interested to learn if these indirect detection observables allow us to glean further detailed information about the dark sector. Dark glueballs are generically long lived particles (LLPs), decaying to the SM on macro length scales. This means if they are produced in colliders they tend to decay outside the detectors. While it is possible to detect the shorter-lived states in LLP searches at the LHC main detectors~\cite{ATLAS:2013bsk,CMS:2018bvr,Alimena:2019zri,ATLAS:2019tkk,CMS:2021dzg} and  proposed external LLP detectors~\cite{MATHUSLA:2018bqv, MATHUSLA:2020uve, Curtin:2018mvb, Aielli:2019ivi, FASER:2018bac, Bauer:2019vqk}, states with extremely long lifetimes may be indetectable due to low production rates, or only show up in missing energy or monojet searches~\cite{Chala:2015ama,ATLAS:2017drc,CMS:2019ysk}, meaning information on the decays themselves is lost. 
Indirect detection searches probe astrophysical length scales and could thus see the entire spectrum of glueball decays, revealing important information about the UV completion of the dark sector and/or the IR nature of pure glue hadronization. 

In this study we compute the observed photon, antiproton and positron spectra in cosmic ray observations for three possible decay portals of glueballs to the SM (Higgs portal, gauge portal, and a Twin-Higgs-like mixture) and several possible benchmarks for the unknown dark hadronization physics. This allows us to provide the first constraints on the annihilation cross section in the parameter space of dark matter and dark glueball mass, and show how this simplified model can fit the Galactic Centre Excess (GCE) \cite{Goodenough:2009gk,Daylan:2014rsa,Calore:2014xka,Cholis:2021rpp,DiMauro:2021raz,Leane:2022bfm}.
We also demonstrate how multi-channel cosmic ray observations can constrain both the unknown details of pure glue hadronization and details of the dark sector UV completion by determining the decay portal for the dark glueballs. 
In particular, under the assumption that dark matter annihilation explains the GCE, decays via the gauge portal are excluded, while the Higgs portal and Twin-Higgs-like case are consistent, especially for $\mathcal{O}(10$ GeV) glueball masses favoured in Fraternal Twin Higgs scenarios~\cite{Craig:2015pha, Curtin:2015fna}.  

Our paper is outlined as follows. In Section. \ref{sec:model} we outline our simplified DM model we utilise in this work. We then go on to explain the various components of the dark glueball shower in Section. \ref{sec:glueballs}, such as the properties and decays of dark glueballs, as well as how we handle pure glue hadronization. Section. \ref{sec:methodology} reviews the various methods we implement to compare our theory to data and put constraints on the DM parameter space. Lastly we provide our results in Sec. \ref{sec:constraints}, the first indirection constraints for dark glueball showers, as well as Sec. \ref{sec:multichannel}, where we outline how our analysis begins to constrains the physics of the dark glueball sector itself. We conclude in Sec. \ref{sec:conclusions}.

\section{Simplified Dark Sector Model}
\label{sec:model}

We consider the following simplified model Lagrangian to represent the generic case of dark matter annihilating to dark gluons which then form dark glueballs:
\begin{equation}
\begin{split}
\mathcal{L} = \overline{q}^D(i D^D - &M_q)q^D - \frac{1}{4} G_{\mu\nu}^D G^{D, \mu\nu} - m_{DM}\overline{\chi}\chi -  m_\phi^2\phi^2 - \lambda_\chi\overline{\chi}\chi\phi -  \lambda_q\overline{q}^D q^D\phi\;.
\end{split}
\end{equation}
We have introduced a fermion, $\chi$, our DM candidate; a dark confining sector containing the dark gluon $G^D$  and a dark quark $q^D$, with mass larger than the confinement scale, $\Lambda$;
and lastly a scalar $\phi$ that couples $\chi$ to the dark gluons via a $q^D$ loop. Minimally a single dark quark is needed for glueballs to decay to the SM, but multiple fields with a range of charge assignments and masses is possible, which we address in Section \ref{subsec:decays}.
A scalar mediator is introduced to couple the DM to the confining sector as for di-gluon production a vector mediator is forbidden by the Landau-Yang theorem for colour-singlet states \cite{Beenakker:2015mra,Cacciari:2015ela}. Note that a vector mediator would still allow dark glueball showers via a three gluon initial state or quirk production. We focus on di-gluon production via a scalar mediator for simplicity.
Thus $\chi$ can annihilate in the galaxy to two dark gluons that then shower and hadronise into dark glueballs. 
Decays of dark glueballs to SM states can be allowed by coupling the dark quarks to the SM Higgs (``Higgs portal'') or making them charged under SM gauge groups (``gauge portal''). We discuss this in more detail in Section \ref{subsec:basics}.

While we make no particular assumption about the DM production mechanism,
a minimal scenario is that $\chi$ is a thermal relic that annihilates to dark gluons in the early universe and makes up all of dark matter today. As we discuss in Section~\ref{sec:constraints}, it is straightforward to reinterpret our results in the more general scenario where $\chi$ is a thermal relic that has other annihilation channels and constitutes only a fraction of DM. This is important since realistic dark sectors can easily feature several types of dark relics, e.g. 
thermal relic abundances of stable glueball states \cite{Forestell:2017wov}  (though this never overcloses the universe for any parameters we consider and  is negligible if the dark sector is cold relative to the standard model \cite{Carenza:2022pjd,Carenza:2023shd}), or 
asymmetric abundances of stable dark quarks in the early universe~\cite{Antipin:2015xia,Mitridate:2017oky}.

Our model is very theory agnostic, only containing the fields needed to produce dark glueball showers. However, it is very much inspired by the Fraternal Twin Higgs (FTH) framework \cite{Craig:2015pha}, in which case our DM candidate $\chi$ would be the mirror tau~\cite{Craig:2015xla, GarciaGarcia:2015fol}, $G_{\mu\nu}^D$ the mirror gluons, $q^D$ the mirror top, and $\phi$ the mirror twin Higgs, or, qualitatively, the twin $Z/\gamma$.\footnote{To be precise, in the FTH model the annihilation of twin tau DM to mirror gluons occurs via twin higgs and $Z$ bosons in the $s$-channel and a dark quark loop to either two or three gluons directly, or to an intermediate quirky~\cite{Kang:2008ea} bound state of mirror bottoms, which then annihilates to mirror gluons. The details of quirky de-excitation and annihilation in this case may be complicated, but the outcome for glueball production should be very similar to the simplified scenario we study, since glueball production should be dominated by s-wave annihilation of the mirror bottoms.}
While the original twin tau DM proposal has since been excluded by direct detection constraints, it was recently understood~\cite{Curtin:2021spx} that including the effects of spontaneous $Z_2$ breaking in the FTH setup makes thermal twin tau dark matter a viable candidate and an attractive target for future direct detection experiments. %

Note that in the case of $q_D$ coupling to both $\phi$ and $h$, scalar mixing will allow $\chi$ to annihilate directly to the SM. However we are interested in the case when this coupling or mixing is small, thus this channel will be suppressed compared to dark gluon production. Small $y_q$ values would also increase the lifetime of the dark glueballs.

\section{Production and Decays of Dark Sector Glueballs}
\label{sec:glueballs}

In this section we briefly summarise the known physics of  dark glueballs. Firstly, we discuss the properties of the various glueball states, such as masses and quantum numbers. We then explain how the dark glueballs are able to decay to the SM via higher dimension operators and how we calculate their branching ratios. Lastly we outline what we know about the production of dark glueballs via pure glue hadronization, its implementation in \texttt{GlueShower}, and importantly how we address uncertainties due to the unknown pure-glue non-perturbative physics.

\subsection{Basics}
\label{subsec:basics}
\begin{figure}[t]
\centering
\includegraphics[width=0.5\linewidth]{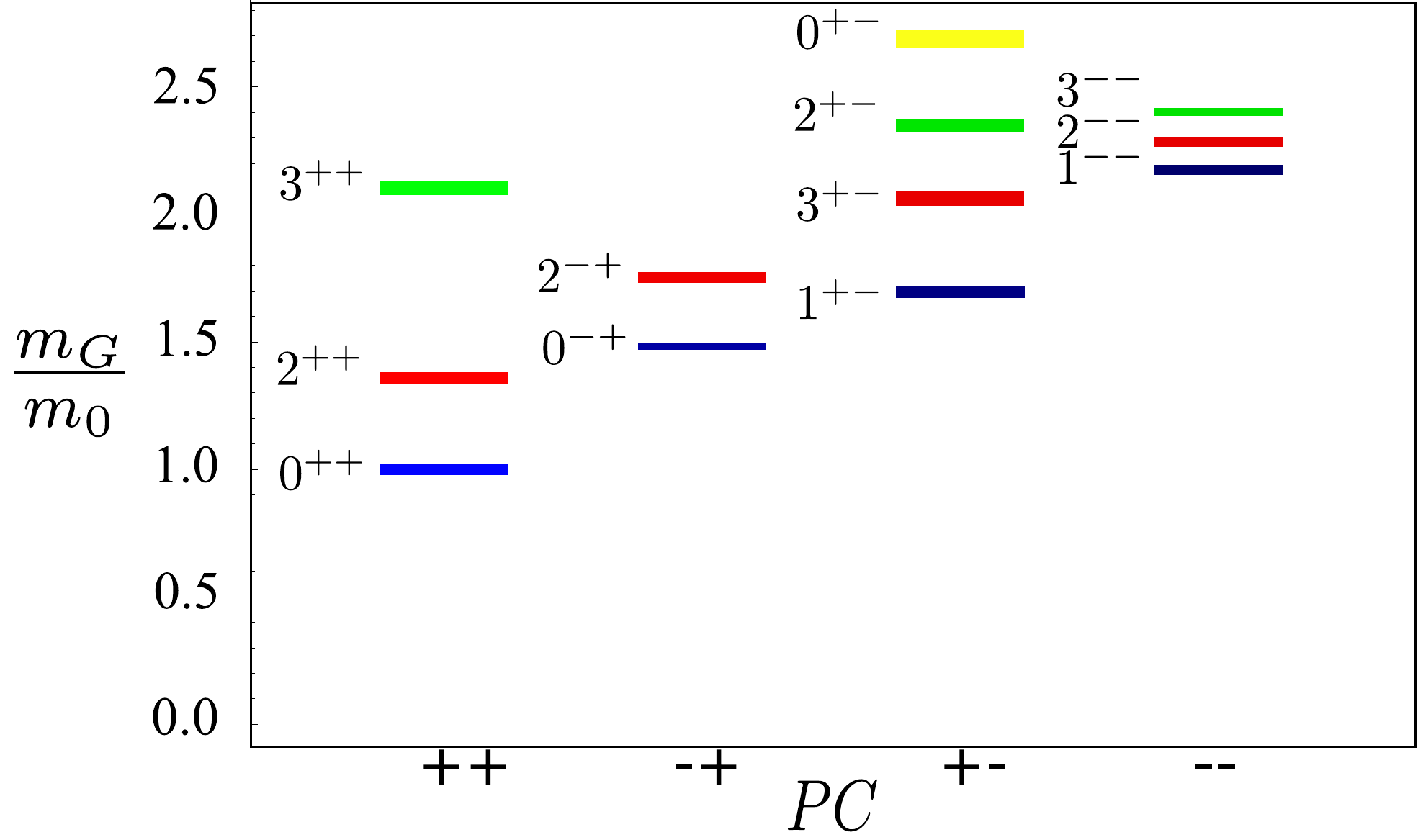}
\caption{Glueball mass $m_G$ spectrum for pure $SU(3)$ Yang-Mills theory \cite{Morningstar:1999rf} in terms of the lightest glueball mass $m_0$, plot taken from \cite{Juknevich:2009gg}.}
\label{fig:glueballspectrum}
\end{figure}

The properties of $SU(N_c)$ glueballs have been studied on the lattice for decades~\cite{Morningstar:1999rf,Teper:1998kw,Lucini:2010nv,Athenodorou:2021qvs,Yamanaka:2021xqh}, establishing a spectrum of twelve stable states in the absence of external couplings, as shown in Figure \ref{fig:glueballspectrum}. The spectrum of masses is given in terms of $m_0$, the lightest glueball mass, and labelled by their quantum numbers $J^{PC}$. Alternatively the masses can be parameterised by the confinement scale of the theory, $\Lambda$, since it has been calculated on the lattice that  $m_0 \approx 6\Lambda$. Thus there is only one free parameter, $m_0$, in determining the entire mass spectrum of glueballs. In this work we will only study the case $N_c = 3$\footnote{Considering different values of $N_c$ would only slightly change the mass ratios for the various glueball states and not significantly change our results.} and use the mass values calculated in Ref. \cite{Athenodorou:2021qvs}, only considering the lightest 10 glueballs. We focus on $m_0$ values in the range 10 - 50 GeV, as this regime is motivated by theories of neutral naturalness \cite{Craig:2015pha,Curtin:2015fna}, but will consider an extended range of glueball masses for completeness. 

Depending on the charge assignment of the dark quark, there are two possible sets of effective operators that couple the dark sector to the SM \cite{Juknevich:2009ji,Juknevich:2009gg}. Firstly, if the dark quark couples to the Higgs field, we get an effective dimension-6 operator:

\begin{equation}
    \mathcal{O}^{(6)} \propto \frac{1}{M^2} H^\dagger H \: \text{tr}\left(G^D G^D\right)
\end{equation}
where $M$ is the mass scale of the dark quarks, $H$ the SM Higgs field, and $G^D$ the dark gluon field strength. In the Twin Higgs model, this arises due to a small required mass mixing between the twin and visible Higgs mass eigenstate. Secondly,  if the dark quark is charged under any of the SM gauge groups, we also get dimension-8 operators:

\begin{align}
    \mathcal{O}^{(8a)} &\propto \frac{1}{M^4} \text{tr}\left(G^{SM} G^{SM}\right) \text{tr}\left(G^D G^D\right)\\
    \mathcal{O}^{(8b)} &\propto \frac{1}{M^4} \text{tr}\left(G^{SM}\right) \text{tr}\left(G^D G^D G^D\right)
\end{align}
where $G^{SM}$ is a SM field strength. In the Twin Higgs model, these operators may arise from a UV completion which often features fields above a TeV charged under both sectors \cite{Geller:2014kta,Barbieri:2015lqa,Low:2015nqa}. These operators allow some or all of the dark glueballs to decay to the SM. Since these operators can be factorised into respective SM and dark sector operators, we need to consider the forms:

\begin{equation}
    \langle SM | \mathcal{O}^{(D-d)}|0\rangle \langle 0 | \mathcal{O}^{(d)} | \Theta_\kappa \rangle
\end{equation}
\begin{equation}
    \langle SM | \mathcal{O}^{(D-d)}|0\rangle \langle \Theta_{\kappa'} | \mathcal{O}^{(d)} | \Theta_\kappa \rangle
\end{equation}
where D is the overall dimension of the operator and $\Theta_\kappa$ refers to a dark glueball with quantum numbers labelled by $\kappa$. The SM part $\langle SM | \mathcal{O}^{(D-d)}|0\rangle$ can be evaluated using perturbative field theory methods, but the dark sector transition matrix elements require non-perturbative physics. We will discuss the specifics of the required matrix elements and decay channels that arise from these operators in the following subsection. 

\subsection{Decay Modes}
\label{subsec:decays}
The relative strengths of the aforementioned Higgs portal and gauge portal operators are effectively a free parameter of our model. 
We analyze three particularly simple or motivated cases:

\begin{itemize}
    \item \textbf{Higgs Portal:} In this regime the dark quark \emph{only} couples to the Higgs, thus decays only proceed through the dimension-6 effective operator. The majority of the dark glueballs are able to decay to the standard model, this is achieved by either directly mixing with the SM Higgs or by decaying to a lighter glueball and radiating an off shell SM Higgs. Note that the $0^{-+}$ and $1^{+-}$ states are entirely stable.
    \item \textbf{Gauge Portal:} In this regime the dark quark \emph{only} couples to the SM gauge fields, thus decays only proceed through the dimension-8 effective operator. The lighter glueballs decay to two SM gauge bosons, while the heavier states mostly decay to a lighter glueball and radiate a single photon, or a $Z$ boson if allowed. In this regime none of the dark glueballs are stable.
    \item \textbf{Twin Higgs-like:} In this regime we assume both operators are active but we introduce two different scales, $M^{(6)}$, the mass scale of the dimension-6 operator, and $M^{(8)}$, the mass scale of the dimension-8 operator. We consider the limit $M^{(6)} \ll M^{(8)}$, inspired by UV completions of Twin Higgs theories \cite{Geller:2014kta,Barbieri:2015lqa,Low:2015nqa}. A mirror top partner, whose mass sets $M^{(6)}$, is present and couples to the SM Higgs but not the SM gauge bosons. At some higher scale, $M^{(8)}$, the theory is UV completed, with fields that have SM gauge charge. In this case due to the mass hierarchy and relative dimensions of the effective operators, dimension-6 decays will heavily dominate over dimension-8 decays. Thus when both decay channels are allowed for a glueball species, we only consider dimension-6 decays, but the previously stable $0^{-+}$ and $1^{+-}$ states will now decay via the dimension-8 operator.
\end{itemize}

\begin{table}
\centering
\resizebox{\textwidth}{!}{
\begin{tabular}{|c|c|c|c|}
\hline
 Glueball & Higgs Portal & Gauge Portal & Twin Higgs-like \\ \hline
 $0^{++}$ & $h^*\rightarrow SM, hh$ & $gg, \gamma\gamma, Z\gamma, ZZ, WW$ & $h^*$ \\ \hline
 $2^{++}$ & $0^{++} + h^*$ & $gg, \gamma\gamma, Z\gamma, ZZ, WW$ & $0^{++} + h^*$ \\ \hline
 $0^{-+}$ & - & $gg, \gamma\gamma, Z\gamma, ZZ, WW$ & $gg, \gamma\gamma, Z\gamma, ZZ, WW$ \\ \hline
 $1^{+-}$ & - & $\{0^{++},2^{++},0^{-+}\} + \{\gamma,Z\}$ & $\{0^{++},2^{++},0^{-+}\} + \gamma$ \\ \hline
 $2^{-+}$ & $0^{-+} + h^*$ & $gg, \gamma\gamma, Z\gamma, ZZ, WW$ & $0^{-+} + h^*$ \\ \hline
 $3^{+-}$ & $1^{+-} + h^*$ & $\{0^{++},2^{++},0^{-+},2^{-+}\} + \{\gamma,Z\}$ & $1^{+-} + h^*$ \\ \hline
 $3^{++}$ & $\{2^{++},0^{-+},2^{-+}\} + h^*$ & $1^{+-} + \{\gamma,Z\}$ & $\{2^{++},0^{-+},2^{-+}\} + h^*$ \\ \hline
 $2^{--}$ & $\{1^{+-},3^{+-}\}+h^*$ & $\{0^{++},2^{++},0^{-+},2^{-+}\} + \{\gamma,Z\}$ & $\{1^{+-},3^{+-}\} + h^*$ \\ \hline
 $1^{--}$ & $1^{+-} + h^*$ & $\{0^{++},2^{++},0^{-+},2^{-+}\} + \{\gamma,Z\}, ff$ & $1^{+-} + h^*$ \\ \hline
 $2^{+-}$ & $\{1^{+-},3^{+-},2^{--},1^{--}\} + h^*$ & $\{0^{++},2^{++},0^{-+},2^{-+}\} + \{\gamma,Z\}$ & $\{1^{+-},3^{+-},2^{--},1^{--}\} + h^*$ \\ \hline
\end{tabular}
}
\caption{Table of decay channels for each glueball we consider. $h^*$ indicates an off shell Higgs, but if allowed will be produced on shell.}
\label{tab:decay_table}
\end{table}

We summarise the decay modes in Table \ref{tab:decay_table}. The specific formula for the various decays can be found in Refs. \cite{Juknevich:2009ji,Juknevich:2009gg} but we will now discuss some of the most important features of this decay phenomenology.

For the Higgs portal case, the $0^{++}$ glueball mixes directly with the Higgs and has branching ratios:

\begin{equation}
    \Gamma_{0^{++}\rightarrow\xi\xi}=\left(\frac{y^2 v \alpha^D_s \mathbf{F^S_{0+}}}{3\pi M^2(m_H^2 - m_0^2)}\right)^2 \Gamma^{SM}_{h\rightarrow\xi\xi}(m_0)
\end{equation}
where $\mathbf{F^S_{0+}} = \langle 0 | \text{tr}(G^D G^D)|0^{++}\rangle$ is the $0^{++}$ annihilation matrix element, and $\Gamma^{SM}_{h\rightarrow\xi\xi}(m_0)$ is the SM Higgs branching ratios as a function of mass, calculated using HDECAY \cite{Djouadi:2018xqq} or chiral perturbation theory~\cite{Winkler:2018qyg} (see below). Thus for the $0^{++}$ state, the branching ratios are equivalent to a SM-like higgs of mass $m_0$. However, at larger $m_0$ values, the $0^{++}\rightarrow hh$ channel also becomes available.

 The rest of the Higgs portal decay modes take the general form, $\Theta_\kappa \rightarrow \Theta_{\kappa '} + h^*$, with the off shell Higgs then decaying to the SM. To calculate the branching ratios for these decays, we need the matrix transition elements, $\langle \Theta_{\kappa'} | \text{tr}(G^D G^D) | \Theta_\kappa \rangle$. While some of these matrix elements have been calculated on the lattice \cite{Chen:2005mg,Meyer:2008tr}, the majority of them remain unknown. However the various matrix elements these studies calculate are within $\mathcal{O}(1)$ factor of one another, so we make the approximation that for the purposes of branching ratios, these matrix elements factorise out and cancel. Under this assumption, the branching ratios are determined solely by the mass splittings between the glueball states. In the $m_0$ range where $h^*$ is produced far off shell, we treat the decay in two steps. Firstly, by binning the two body phase space as a function of $m_{h^*}$, we are able to define branching ratios to $h^*$ of various mass values. Once the glueball has decayed to the off shell Higgs with a specific mass value, we use HDECAY (if $m_{h*} > 2$ GeV) or chiral perturbation theory~\cite{Winkler:2018qyg} (if $m_{h*} < 2$ GeV) to determine its branching ratios into SM fundamental particles or hadrons. In the case that the Higgs can be produced on shell, it is treated as a simple two body decay.

For dimension-8 operators, the branching ratios to different gauge groups is dependent on the representation(s) of the heavy field(s). This is encoded in the various coefficients that appear in the full Lagrangian \cite{Juknevich:2009gg,Juknevich:2009ji,Forestell:2017wov}:

\begin{align}
    \chi_i &= \sum_r d(r_i) T_2(r_i)/\rho_r^4 \\
    \chi_Y &= \sum_r d(r_i) Y_r/\rho_r^4
\end{align}
where $r$ is the SM representation of the heavy dark quarks with masses $M_r$, and $\rho_r = M_r/M$. Index, $i$, runs over the sub-representations corresponding to the SM gauge groups:  $U(1)_Y,SU(2)_L,SU(3)_c$. $d(r_i)$ indicates the number of copies of the $i$-th subrepresentation, and $T_2(r_i)$ is the trace invariant for that factor. To remain relatively theory agnostic about the UV completion of the model, we simply set $\chi_i = \chi_Y = 1$. This gives us the relation:

\begin{equation}
    \frac{\Theta\rightarrow\gamma\gamma}{\Theta\rightarrow g g} = \frac{1}{2}\left(\frac{\alpha}{\alpha_s}\right)^2,
\end{equation}
which applies for $\Theta = 0^{++}, 2^{++}, 0^{-+}, 2^{-+}$. For small $m_0$ values, these are the only significant decay channels for the dark glueballs listed. Thus the dimension-8 branching ratios for the lighter glueballs are entirely determined by SM parameters. As $m_0$ increases, the $Z\gamma,ZZ,WW$ channels become available and their branching ratios can be found in Ref.~\cite{Juknevich:2009gg,Juknevich:2009ji}. 

The majority of the heavier glueball dimension-8 decays are of the form, $\Theta_\kappa \rightarrow \Theta_{\kappa '} + \{\gamma,Z\}$. Once we assume the corresponding matrix elements are of the same order and cancel, the branching ratios are determined by the glueball mass splittings and a spin dependent factor.

The $3^{++}$ state is noteworthy as it has three competing decay modes due to the dimension 8 operator, and a large uncertainty regarding which dominate. In this work we simply assume $3^{++} \rightarrow 1^{+-} + \{\gamma,Z\}$, ignoring the $3^{++} \rightarrow \{2^{++},0^{-+}\} + \{\gamma,Z\}$ decay modes. We select this decay mode as it is the most distinct and will provide the most conservative constraints. Additionally this will be a small effect on the total spectrum. The $3^{++}$ state is a heavy glueball and produced in relatively small amounts compared to the lighter glueballs across the hadronization benchmarks, as we explain in the following subsection.

\subsection{Pure Glue Hadronization}
\label{subsec:glueshower}
To calculate the energy spectra of the SM particles produced in dark glueball showers, we first need the energy spectra of the various dark glueball states produced in DM annihilation. These spectra, also known as fragmention functions, are dependent on the model used to shower and hadronize the initial dark gluons into the final state dark glueballs. In this work we will use \texttt{GlueShower}\footnote{\href{https://github.com/davidrcurtin/glueshower}{https://github.com/davidrcurtin/glueshower}}~\cite{Curtin:2022tou}  to generate these dark glueball fragmentation functions.

We will briefly summarise the underlying physics \texttt{GlueShower} utilises in its algorithm, see Ref.~\cite{Curtin:2022tou} for a more in-depth discussion. Firstly, like any $SU(N_c)$ theory, the shower evolution of the initial virtual dark gluons is determined by perturbative QCD. This allows us to calculate the virtualities and multiplicity of dark gluons as the shower evolves towards the confinement scale. Assuming that $N_f = 0$ QCD satisfies the assumption that hadronization is an IR process active around the confinement scale (referred to as the `jet-like' case), the shower is then terminated not at the confinement scale but at $\Lambda_{had} \geq 2  m_0$.
This is due to the fact perturbative estimates of momentum exchange between separate branches of the shower past this point were found to be very small, and therefore such a dark gluon can no longer split and form two on shell glueballs once confinement occurs.
The dark gluon is hence simply put on shell as a single dark glueball.

As the non-perturbative physics of pure glue hadronization are a priori unknown, \texttt{GlueShower} also allows an alternative shower possibility in which fragmentation occurs at large scales relative to the confinement scale (referred to as the `plasma-like' case). Once the perturbative shower has ended, the virtual gluons are treated like a pure glue plasma. These plasma states then de-excite by quasi-isotropically emitting on shell glueballs with thermal momenta, similar to the evaporation of quark gluon plasma by pion emission. In general one observes that `plasma-like' showers result in higher multiplicity showers with softer glueballs, compared to the `jet-like' case. 

In both cases described above, the relative probabilities of the various glueball states are determined by a Boltzmann distribution, controlled by the hadronization temperature, $T_{had}$, calculated from lattice QCD \cite{Lucini:2012wq}. Thus the production of heavier glueball states are exponentially suppressed. 

\begin{table}
\centering
\begin{tabular}{|c|c|c|c|}
\hline
 Hadronization  & Hadronization  & Hadronization & Shower \\
 Benchmark & Scale & Temperature & Type \\
 Label & $\Lambda_{had}/2m_0$ & $T_{had}/T_c$ &  \\ \hline \hline
 $J_{1,1}$ & 1 & 1 & \texttt{Jet-like} \\ \hline
 $J_{1,2}$ & 1 & 2 & \texttt{Jet-like} \\ \hline
 $J_{2,1}$ & 2 & 1 & \texttt{Jet-like} \\ \hline
 $J_{2,2}$ & 2 & 2 & \texttt{Jet-like} \\ \hline \hline
 $P_{4,1}$ & 4 & 1 & \texttt{Plasma-like} \\ \hline
 $P_{4,2}$ & 4 & 2 & \texttt{Plasma-like} \\ \hline
 $P_{6,1}$ & 6 & 1 & \texttt{Plasma-like} \\ \hline
 $P_{6,2}$ & 6 & 2 & \texttt{Plasma-like} \\ \hline
\end{tabular}
\caption{Pure-glue hadronization benchmarks.}
\label{tab:hadronization_table}
\end{table}

\begin{figure}[t]
\includegraphics[width=\textwidth]{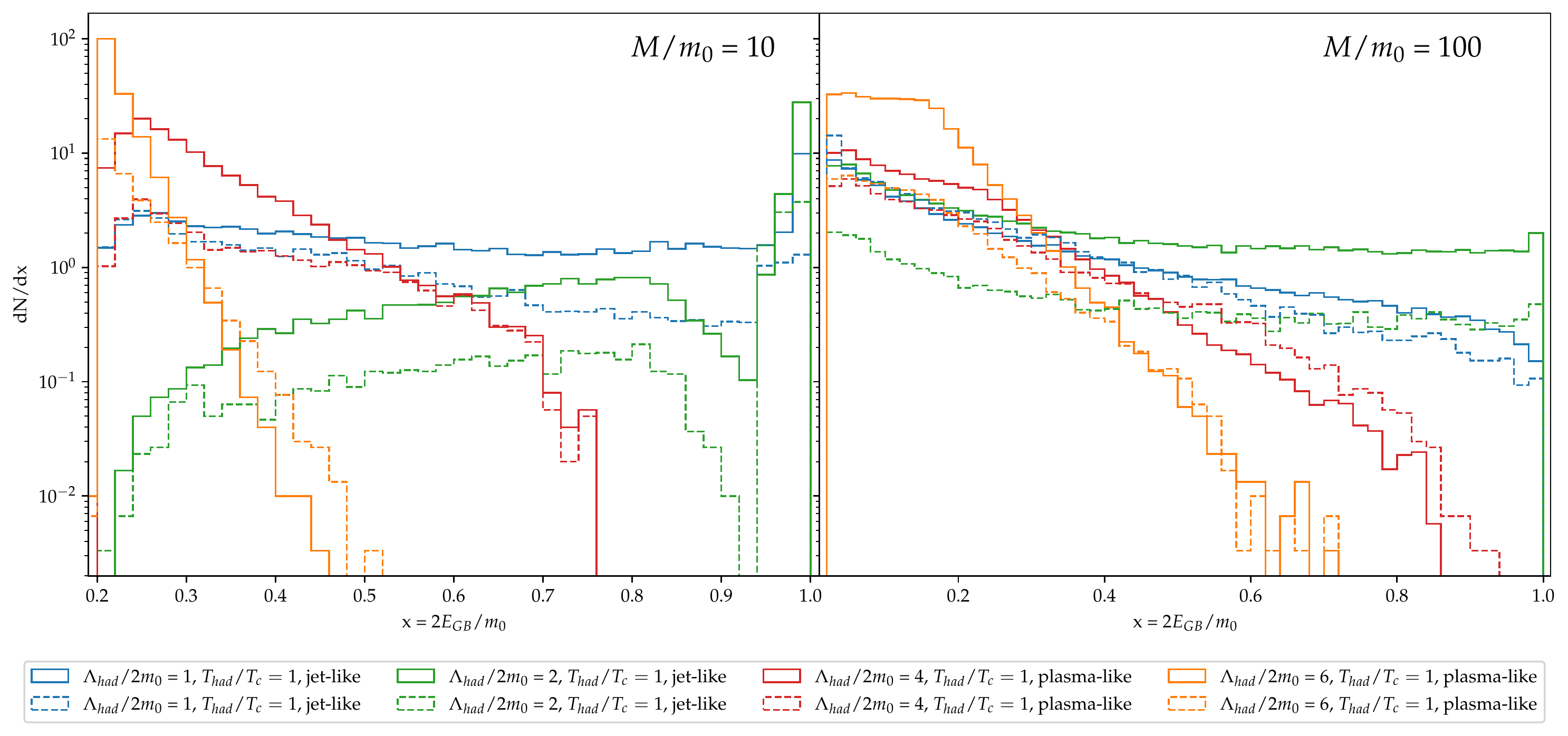}
\caption{$0^{++}$ glueball fragmentation functions for $M/m_0$ = 10 (left) and 100 (right) and $N_c$ = 3, for our eight hadronization benchmarks.}
\label{fig:benchmark_FF}
\end{figure}

In addition to the qualitative choice of shower behaviour, two nuisance parameters are included to account for the unknown quantitative details of pure-glue hadronization:
$\Lambda_{had}/2 m_0 \geq 1$ affects 
the final state multiplicity of dark glueballs, and $T_{had}/T_c \geq 1$ determines the relative production of heavier vs lighter dark glueball states. The dependence of glueball fragmentation functions on these nuisance parameters is explored in Ref. \cite{Curtin:2022tou} and they provide a set of eight benchmark parameter choices that bracket the possible range of phenomena. In Table. \ref{tab:hadronization_table} we list the eight benchmark sets of nuissance parameters and an example of the resultant fragmentation functions corresponding to these benchmarks are shown in Fig. \ref{fig:benchmark_FF}. The range in behaviour of these fragmentation functions is used in our study to incorporate the theoretical hadronization uncertainties into our final constraints.

An interesting limit to consider is when the centre of mass energy of the shower $E$ is less than the scale at which we end the perturbative shower, $2m_0 \leq E \leq \Lambda_{had}$. For the jet-like case, each gluon is not allowed to perturbatively split as its virtuality is already below the hadronization scale. Thus for $E \leq \Lambda_{had}$, jet-like showers always result in back-to-back glueball production. For the plasma-like case, again the gluons are unable to perturbatively split. However, in this case the gluons are then treated as a single plasma at rest that de-excites quasi-isotropically as mentioned before.

\section{Methodology}
\label{sec:methodology}

To study DM annihilating into dark glueballs, we first must calculate the resulting SM spectra in the galactic rest frame. We show this for a variety of dark matter and glueball masses, as well as across our three decay portals and eight hadronization benchmarks. Once an assumption is made regarding the distribution of DM in our galaxy, we can then calculate the observed flux at Earth's position. For charged particles this requires cosmic ray propagation and solar modulation tools. Finally with the flux at Earth calculated, we are able to find constraints by comparing the spectra to observational data by Fermi-LAT in the case of photons, and AMS-02 for antiprotons and positrons.

\subsection{Monte Carlo Generation of SM final state spectra in DM rest frame}
\label{subsec:MCspectra}

For each species of dark glueball, we generate the energy spectra of photons, positrons, and antiprotons from their decays using PYTHIA 8 \cite{Bierlich:2022pfr}. The fragmentation function for each dark glueball species, produced in the annihilation of DM to dark gluons, is generated using \texttt{GlueShower}. We then Lorentz boost and convolve the SM spectra with the dark glueball fragmentation functions to get the SM energy spectra for the entire dark glueball shower in the galactic frame.

\begin{figure}[t]
\includegraphics[width=\textwidth]{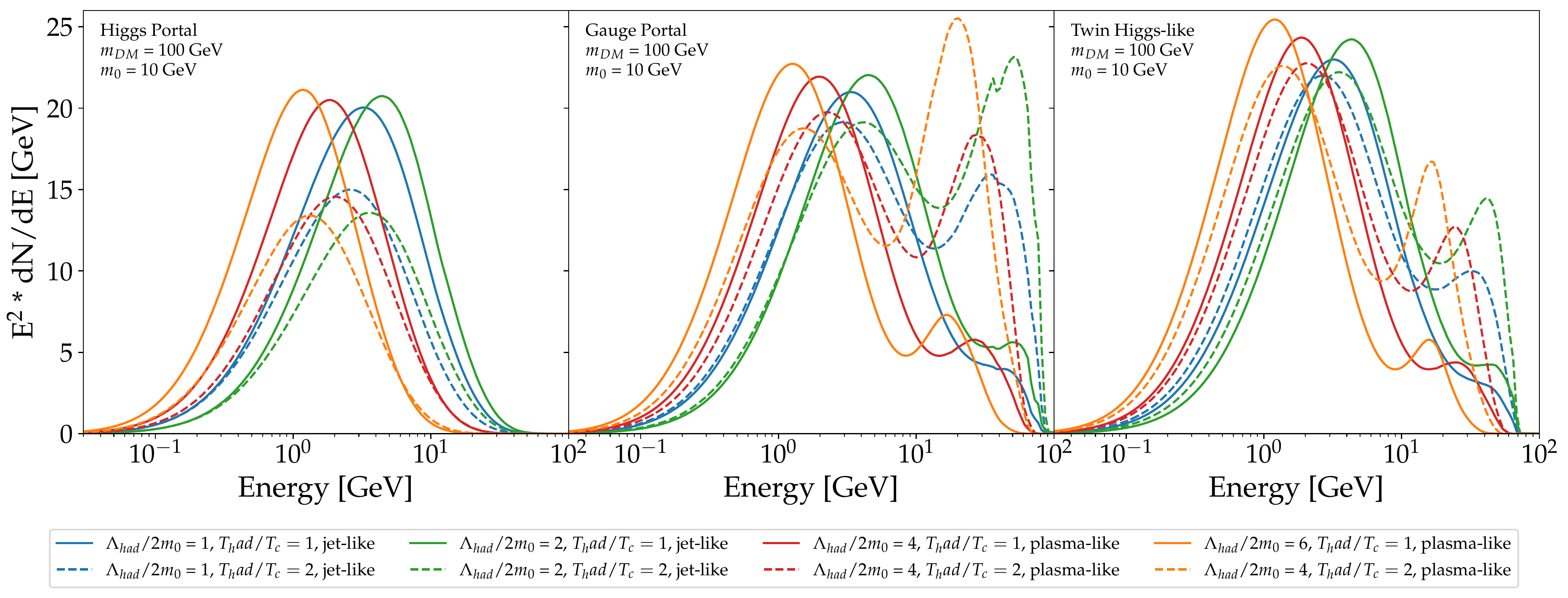}
\caption{Photon spectra in the galactic rest frame for $m_{DM} = 100$ GeV and $m_0 = 10$ GeV. Depicted are the spectra for each of the eight hadronization benchmarks and three decay portal cases we consider.}
\label{fig:photon_spectra}
\end{figure}

Fig. \ref{fig:photon_spectra} depicts the photon spectra in the galactic frame for a reference dark matter and glueball mass. In the simplest Higgs portal case we see how our assumptions about hadronization affect the spectral energy peak position as well as total flux. Plasma-like showers lead to softer spectra compared to jet-like showers, and how higher temperatures lead to overall smaller multiplicity due to an increased production of heavier, stable glueballs. In both the gauge portal and Twin Higgs-like cases we see the effect of heavier glueballs decaying to lighter glueballs and a single photon. Due to the range of produced glueball energies, this spectral line is smoothed out to a second, high energy peak in the spectra. The presence and height of this second peak is largely influenced by the hadronization temperature.

\begin{figure}[t]
\includegraphics[width=\textwidth]{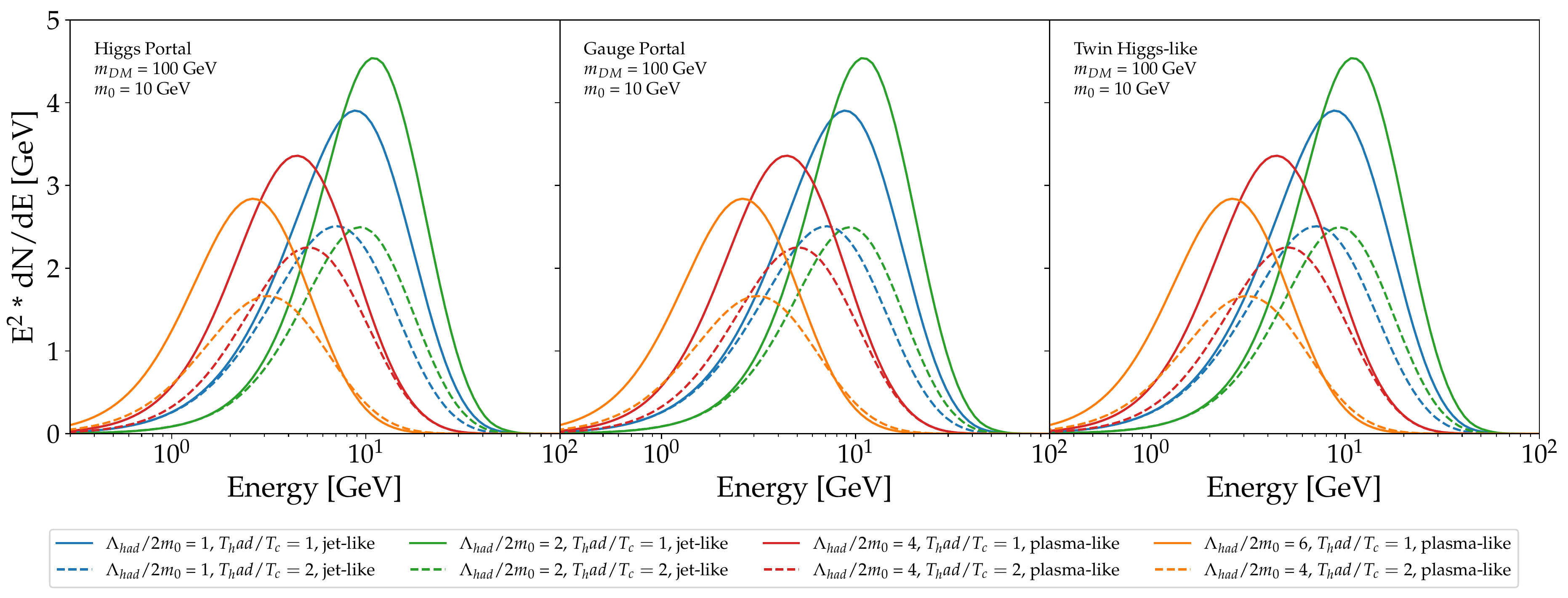}
\caption{Antiproton spectra in the galactic rest frame for $m_{DM} = 100$ GeV and $m_0 = 10$ GeV. Depicted are the spectra for each of the eight hadronization benchmarks and three decay portal cases we consider.}
\label{fig:antiproton_spectra}
\end{figure}

We also show the antiproton spectra, in the galactic frame, in Fig. \ref{fig:antiproton_spectra}. We find there is less of a dependence on the decay portal determining the spectral shape, compared to the photon channel. However, there are larger differences in peak positions and total flux depending on the hadronization assumption used for the antiprotons. This is to be expected as the antiproton's energy is more influenced by the boost provided by the parent glueball due to its large mass compared to the photon. Thus the antiproton channel is more sensitive to the glueball fragmentation function and therefore hadronization assumptions.

Once we have calculated the SM spectra in the galactic frame, to get the actual observed flux at Earth, we must make an assumption regarding the distribution of DM in the galaxy, which we will describe in the following section. Additionally for positrons and antiprotons, their galactic frame spectra have to then be propagated to the solar system's position using a cosmic ray propagation tool. This is due to their trajectories being influenced by the galactic magnetic fiields. Commonly used examples are fully numerical codes such as GALPROP \cite{Strong:1998fr}, DRAGON \cite{Evoli:2008dv,Evoli:2016xgn}, and PICARD \cite{Kissmann:2014sia}, or even semi-analytical tools such as USINE \cite{Maurin:2018rmm}. Once the positron and antiproton spectra have been calculated at the solar system's position, we then consider solar modulation effects from the heliosphere. We outline our method of cosmic ray propagation and solar modulation in Sec \ref{subsec:CRpropagation}.

\subsection{Cosmic Ray Propagation}
\label{subsec:CRpropagation}

The evolution of the number density spectra, $N_i(\vec{r},p)$, throughout the galaxy for each particle species we consider is given by the diffusion equation \cite{1964ocr..book.....G,1990acr..book.....B},

\begin{multline}
\vec{\nabla} \cdot \left(D(R)\vec{\nabla} - \vec{v_\omega}\right)N_i + \frac{\partial}{\partial p}\left[p^2 D_{pp}(R)\frac{\partial}{\partial p}\left(\frac{N_i}{p^2}\right)\right] - \frac{\partial}{\partial p}\left[\dot{p}N_i - \frac{p}{3}\left(\vec{\nabla}\cdot \vec{v_\omega}\right)N_i\right] = \\ Q_i + \sum_{i<j}\left(c\beta n_{gas}\sigma_{j\rightarrow i} + \frac{1}{\gamma \tau_{j\rightarrow i}} \right)N_j - \left(c\beta n_{gas}\sigma_i + \frac{1}{\gamma \tau_i} \right)N_i
\end{multline}
where $D(R)$ and $D_{pp}(R)$ are the spatial and momentum diffusion coefficients respectively. Apart from spatial diffusion and momentum diffusion (also referred to as reacceleration), this equation also includes the effects of convection, energy losses, injection from sources, and decays and collisions with the interstellar medium.

In this work we use a customised version of the DRAGON2 code\footnote{\href{https://github.com/tospines/Customised-DRAGON2_beta}{https://github.com/tospines/Customised-DRAGON2\_beta}} to solve the CR diffusion equation and calculate propagated energy spectra for various particle species. This code uses a spatial diffusion coefficient with a rigidity dependent break, taking the form

\begin{equation}
    D(R) = D_0 \beta^\eta \left(\frac{R}{R_0}\right)^\delta\left[1 + \left(\frac{R}{R_b}\right)^{\Delta\delta / s}\right]^{-s}
\end{equation}
where $R_0 = 4\:GV$. Recent studies have shown that CR data favours a diffusion coefficient with a high-rigidity break \cite{Genolini:2017dfb}, and from this work we use their parameterisation with $R_b = 312\:GV$, $\Delta\delta = 0.14$, $s = 0.04$.

The momentum diffusion coefficient takes the form

\begin{equation}
    D_{pp}(R) = \frac{4}{3} \frac{1}{\delta(4 - \delta^2)(4 - \delta)} \frac{V_A^2 p^2}{D(R)}
\end{equation}
which is related to the spatial diffusion coefficient, as well as the Alfv\'en velocity, $V_A$. For the remaining free parameters we implement the values found in Ref. \cite{Luque:2021ddh}, which also used the customised DRAGON2 code and found these values by performing a combined analysis, fitting multiple simulated CR ratios and spectra to observed cosmic ray data.

The DM source contribution is given by:

\begin{equation}
    Q_\chi(\vec{\bm{r}}) = \frac{\rho(\vec{\bm{r}})^2 \langle \sigma v \rangle}{2 m_\chi^2}\frac{dN}{dE}
\end{equation}
where $\rho(\vec{\bm{r}})$ is the DM halo density profile, $\langle \sigma v \rangle$ is the thermally averaged interaction cross section, and $dN/dE$ is the energy spectra from DM annihilation.
To account for solar modulation once the spectra have been propagated with DRAGON2, we use the force-field approximation \cite{1976ApJ...206..333F}, characterised by the Fisk potential, $\phi$:

\begin{equation}
    \frac{dN^\oplus}{dE_{kin}}(E_{kin}) = \frac{(E_{kin} + m)^2 - m^2}{(E_{kin} + m + |Z|e\phi)^2 - m^2} \frac{dN^{ISM}}{dE_{kin}}(E_{kin}  |Z|e\phi)
\end{equation}

However, solar modulation is known to be both time and charge dependent, and attempts have been made to use more sophisticated solar modulation techniques. To incorporate some of these effects we use a charge-dependent Fisk potential:

\begin{equation}
    \phi = \phi_0 + \phi_1^{\pm}\left(\frac{1\:GV}{R}\right)
\end{equation}
where $\phi_1^{\pm}$ is a charge-dependent term that accounts for increased energy loss when a particle's sign opposes the solar polarity. Thus we assume $\phi_1^{+} = 0$, and let $\phi_1^{-}$ be non-zero. While this is slightly more complex than the basic force-field approach, we note this is still an approximation and there exist many more involved methods where the solar propagation is solved numerically \cite{Kappl:2015hxv,Vittino:2017fuh,Boschini:2017gic}.

\subsection{Comparison to Data}
\label{subsec:data}
For the dark glueball decay chains we consider in this work, the majority of SM states that are produced are hadronic, arising from the production of $b\overline{b}$ or $gg$ pairs in dark glueball decays. Searches for antiprotons and gamma-rays produced by these events provide the best option for finding a potential signal but also put the strongest constraints on the DM parameter space. Additionally we also include constraints from positron searches for the sake of completeness and as an extra search channel, even though they provide less competitive constraints. In the following sections we outline how these constraints are calculated from the observed data.

\subsubsection{Galactic Centre Excess}
\label{GCE}

Various interpretations of the status of the GCE exist in the literature\footnote{See Ref. \cite{Leane:2022bfm} for a comprehensive review}, with different analyses changing the shape of the perceived excess \cite{Calore:2014xka,Cholis:2021rpp,DiMauro:2021raz}. We fit our DM flux, given by:

\begin{equation}
    \Phi = \frac{\langle\sigma v\rangle}{8\pi m_{DM}^2} \left [ \int_{E_{min}}^{E_{max}} E \frac{dN}{dE} dE \right ] J_{GCE}
\end{equation}
to the excess as calculated in Ref. \cite{DiMauro:2021raz}. The value of $J_{GCE}$ is determined by both the region of interest and DM density profile. Following the analysis of Ref. \cite{DiMauro:2021qcf}, we assume a ROI of $40\degree \times 40\degree$ centred on the galactic centre, and take their \texttt{MED} DM density profile parameterisation. They use a generalised Navarro-Frenk-White (NFW) function \cite{Navarro:1996gj}, of the form:

\begin{equation}
    \rho_{gNFW} = \frac{\rho_s}{\left(\frac{r}{r_s}\right)^\gamma \left(1 + \frac{r}{r_s}\right)^{3 - \gamma}}
\end{equation}
where $\gamma=1.3$, $\rho_s= 0.449$ GeV cm$^{-3}$, and $r_s = 12.67$ kpc. This DM density profile is also used for the DM source term when using DRAGON2 to propagate the antiproton and positron spectra.

\subsubsection{Fermi-LAT Dwarf Spheroidal Gamma-ray Limits}
\label{subsubsec:fermi}
We follow the official Fermi analysis on Pass 8 LAT data \cite{Fermi-LAT:2016uux} to find constraints on DM annihilating to photons via a dark glueball shower. This method is a combined analysis of 41 galaxies, 28 of which are kinematically confirmed, while the remaining 13 are considered likely galaxies.

Fermi publicly releases their bin-by-bin likelihood functions\footnote{\href{http://www-glast.stanford.edu/pub data/1203/}{http://www-glast.stanford.edu/pub data/1203/}}, as a function of energy flux for each dwarf galaxy. The uncertainty in the J-factor is also included in our likelihood function as a nuisance parameter we profile over, following the profile likelihood method \cite{Rolke:2004mj}. This modifies the likelihood for each galaxy in the following manner:

\begin{equation}
    \Tilde{\mathcal{L}_i}(\mu,J_i | \mathcal{D}_i) = \mathcal{L}_i(\mu | \mathcal{D}_i) \times \frac{1}{\text{ln}(10)J_{\text{obs},i}\sqrt{2\pi}\sigma_i}\:\text{exp}\left[ \frac{-(\text{log}_{10}(J_i) - \text{log}_{10}(J_{\text{obs},i}))^2}{2\sigma_i^2} \right ]
\end{equation}
where $J_i$ is the true value of J-factor, while $J_{\text{obs},i}$ is the measured J-factor with error $\sigma_i$. Values for $J_{\text{obs},i}$ and $\sigma_i$ are taken from Ref. \cite{Fermi-LAT:2016uux} when provided, but in the case that the data is not available  we use the predicted value, $J_{\text{pred},i}$, with a nominal error of 0.6 dex, assuming a NFW profile. Constraints on $\langle \sigma v \rangle $ for a given $m_{DM}$ are calculated using the delta-log-likelihood technique, which for a 95\% C.L upper limit requires a change of 2.71/2 from the maximum value.

\subsubsection{AMS-02 Antiprotons}
\label{subsubsec:antiprotons}

We compare our generated spectra to the latest antiproton spectra released by AMS-02 \cite{AMS:2021nhj}. We assume the data observed by AMS-02 contains no DM source contributions, i.e. the spectra is solely SM background. While there has previously been claims of an antiproton excess that could be explained by a DM component \cite{Cuoco:2016eej,Cui:2016ppb}, establishing the significance of these claims has been difficult since the error correlation matrix of the AMS-02 analysis is not publicly available. Recent analyses using the updated data we consider in this work \cite{DiMauro:2021qcf,Kahlhoefer:2021sha}, or implementing their own error covariance estimates \cite{Boudaud:2019efq,Heisig:2020nse}, were able to reduce the significance of the signal or even make it go away. It is from these developments we make the previous assumption that the observed data is solely SM background.

From this assumption we choose to keep the DRAGON2 propagation parameters fixed for all DM parameter choices as they best fit the observed SM data. Due to the uncertainty in solar modulation and antiproton cross section, we let the solar modulation parameters and SM antiproton spectra normalisation, $\mathcal{N}$, float along with the DM signal normalisation to find the best fit. From combined fits to the AMS-02 \cite{PhysRevLett.114.171103} and Voyager proton data \cite{doi:10.1126/science.1236408}, the Fisk potential parameter was determined to lie within the range, $\phi_0 = 0.60 - 0.72$, which we let it float within. $\phi_1^{-}$ is allowed to float freely. There is roughly 10\% uncertainty on the secondary $\overline{p}$ cross section \cite{Winkler:2017xor}, important to the SM background processes where $\overline{p}$ are produced by protons scattering off the interstella medium. Due to this uncertainty we let $\mathcal{N}$ float in the range 0.9-1.1.

Additionally, to minimise the  influence of assumptions made about solar modulation, which is known to only have large effects on low-energy cosmic rays, we only fit the spectra above 4 GeV. The best fit is determined by minimising:

\begin{equation}
    \chi^2(\theta) = \sum_i \left(\frac{\Phi_i^{theory}(\theta) - \Phi_i^{data}}{\sigma_i}\right)^2
\end{equation}
where $\theta$ is the set of our theory parameters. The DM signal normalisation is then increased until the fit worsens such that the $\chi^2$ value increases by 2.71, corresponding to a 95\% C.L. upper limit on $\langle \sigma v \rangle$.

\subsubsection{AMS-02 Positrons}
\label{subsubsec:positrons}
Similar in process to the antiproton constraints, we use the latest AMS-02 positron data \cite{PhysRevLett.122.041102} and assume it contains no DM contributions. However, for the positron spectra AMS-02 provides an analytical function that provides a good fit to the data. We use this function as our SM background and add a DM contribution found using DRAGON2. While we fit the DM contribution to the observed data, the analytic function's free parameters are allowed to float within their significance range. Again the quality of the fit is determined by  $\chi^2(\theta)$ and a 95\% C.L. upper limit on $\langle \sigma v \rangle$ is found by increasing the DM signal normalisation until $\chi^2(\theta)$ increases by  2.71.

\section{Constraints on Dark Matter annihilating to Dark Glueball Showers}
\label{sec:constraints}

Here we present the 95\% confidence limits on DM $\langle\sigma v \rangle$ across photon, antiproton, and positron channels. Photon constraints are found by comparing to Fermi-LAT dwarf spheroidal galaxy observations, while both antiproton and positron constraints are found by comparing to AMS-02 cosmic ray data. To incorporate the theoretical uncertainty of our hadronization models used, we calculate the confidence limit for each of the eight hadronization benchmarks. Treating the jet-like and plasma-like cases separately, due to their largely different underlying assumptions, we then depict a single constraint limit for each case, with a width defined by the maximum and minimum values across the respective four hadronization benchmarks we consider.

In the following plots we include the thermal relic cross section $\langle\sigma v \rangle_{thermal}$ \cite{Steigman:2012nb} as a benchmark. 
We show excluded annihilation cross sections, $\langle\sigma v \rangle_{limit}$, which are computed assuming that $\chi$ makes up all of the DM and annihilates exclusively into dark gluons, but our results can be interpreted more generally. 
If the sensitivity exceeds the thermal benchmark $\langle\sigma v \rangle_{limit} <  \langle\sigma v \rangle_{thermal}$, then this effectively constraints

\begin{equation}
    \Bigg(\frac{\rho_\chi}{\rho_{DM}}\Bigg)^2 \ \times \ \Bigg( \frac{\langle\sigma v \rangle_{\chi \chi \to G^D G^D}}{\langle\sigma v \rangle_{thermal}}\Bigg) \ < \ \frac{\langle\sigma v \rangle_{limit}}{\langle\sigma v \rangle_{thermal}} \ , 
\end{equation}
in the more general scenario where $\chi$ may only constitute some mass fraction of DM and may only produce dark glueballs in a fraction of its annihilations. 
On the other hand, $\langle\sigma v \rangle_{limit} > \langle\sigma v \rangle_{thermal}$ for a given set of parameters means we are not sensitive to any thermal relic $\chi$ regardless of its DM mass fraction or annihilation fraction to dark glueballs.



\subsection{Fermi-LAT Dwarf Spheroidal Galaxies}
\label{subsec:fermi}

\begin{figure}[t]
\begin{tabular}{cc}
\includegraphics[width=\textwidth]{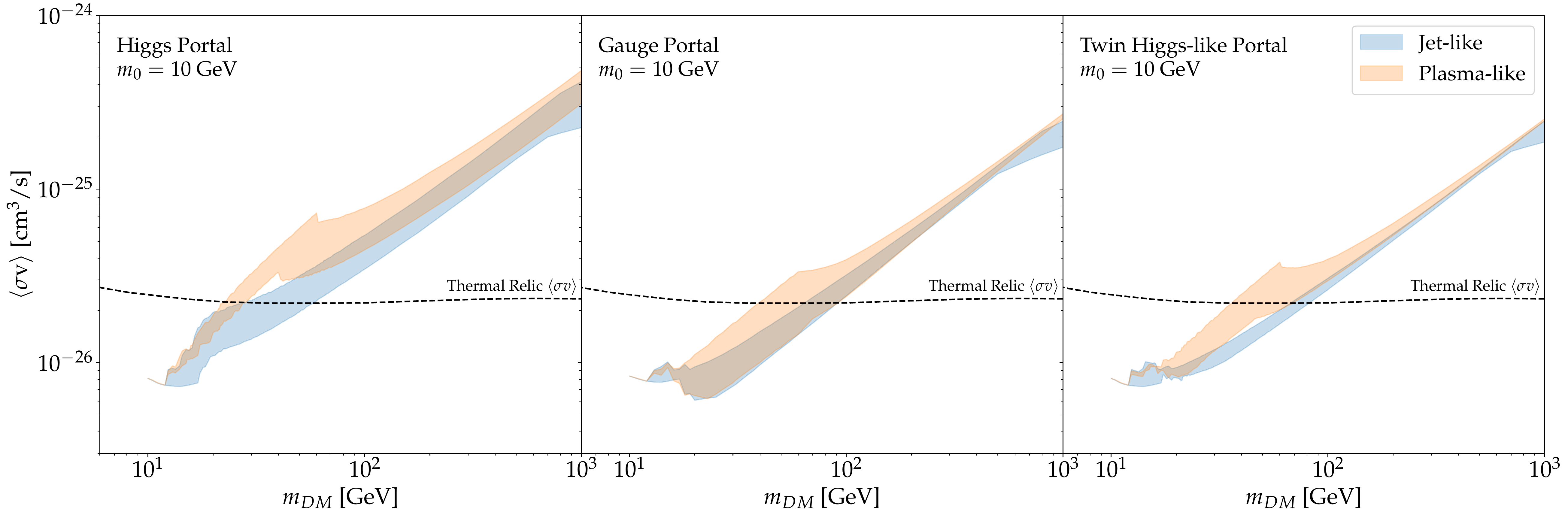}
\\
\includegraphics[width=\textwidth]{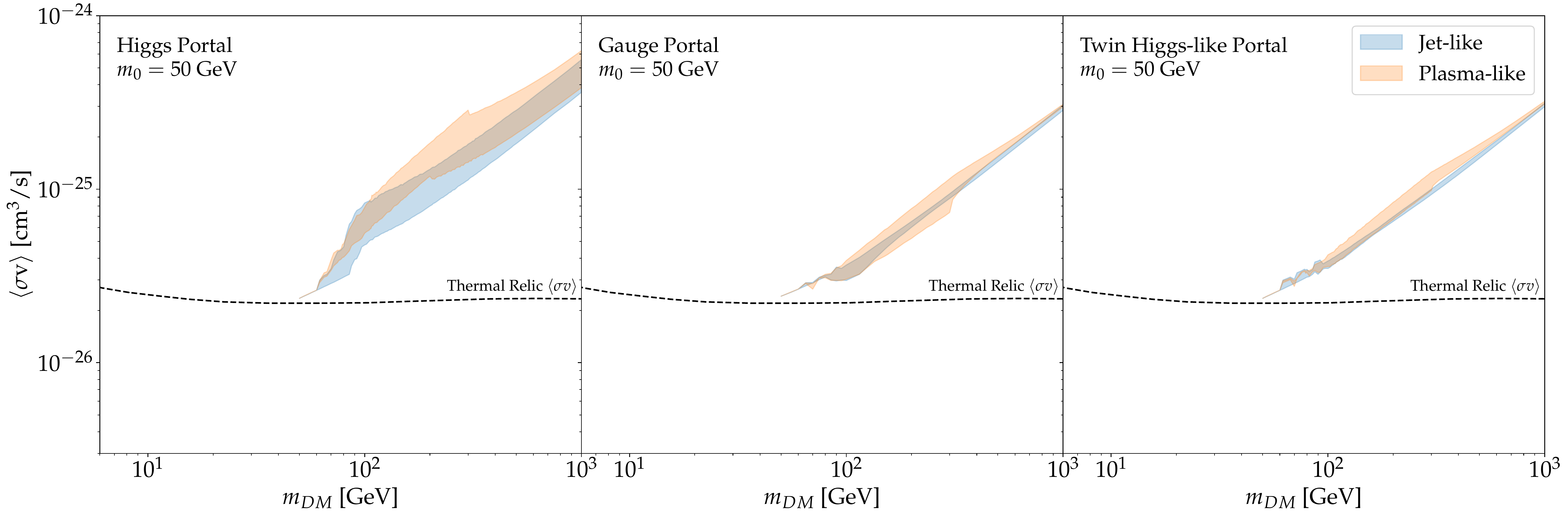}
\end{tabular}
\caption{Fermi-LAT Dwarf Spheroidal limits at 95\% C.L. for DM annihilation to dark glueball showers with $m_0 = 10$ GeV (top) and 50 GeV (bottom). We plot the limits for jet-like (blue) and plasma-like (orange) showers separately for each of the decay portals we consider. The thermal relic cross section is the black dashed line \cite{Steigman:2012nb}.}
\label{fig:constraints_photon_10GeV}
\end{figure}

\begin{figure}[t]
\begin{tabular}{c}
\includegraphics[width=\textwidth]{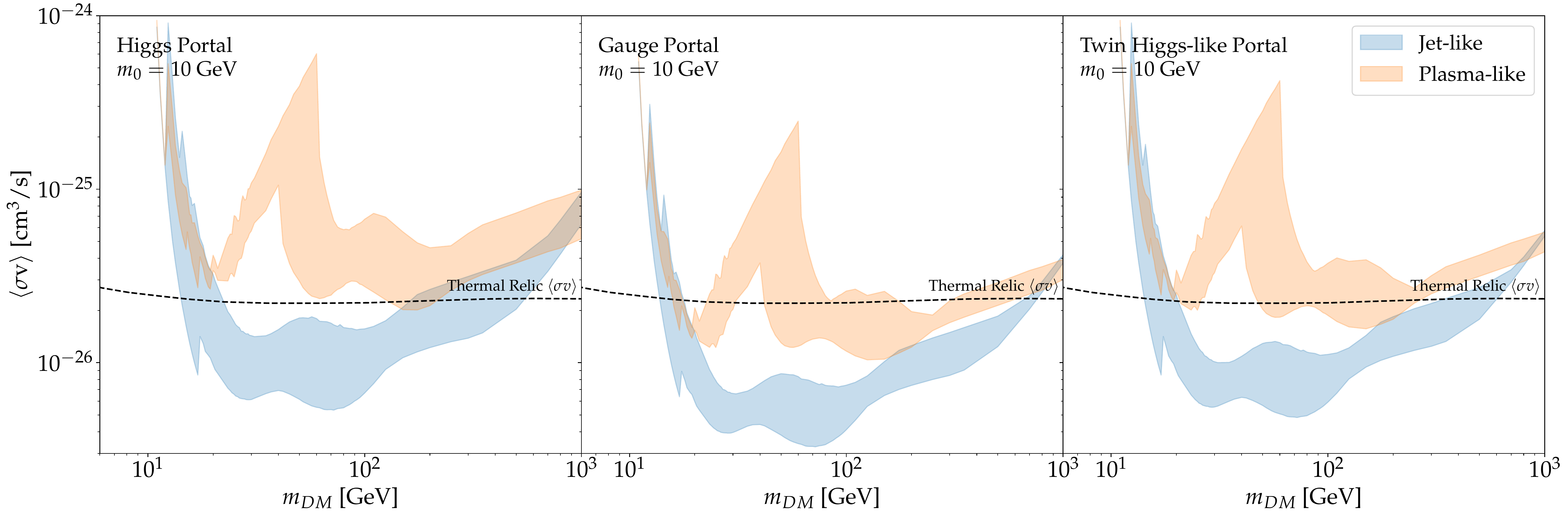}
\\
\includegraphics[width=\textwidth]{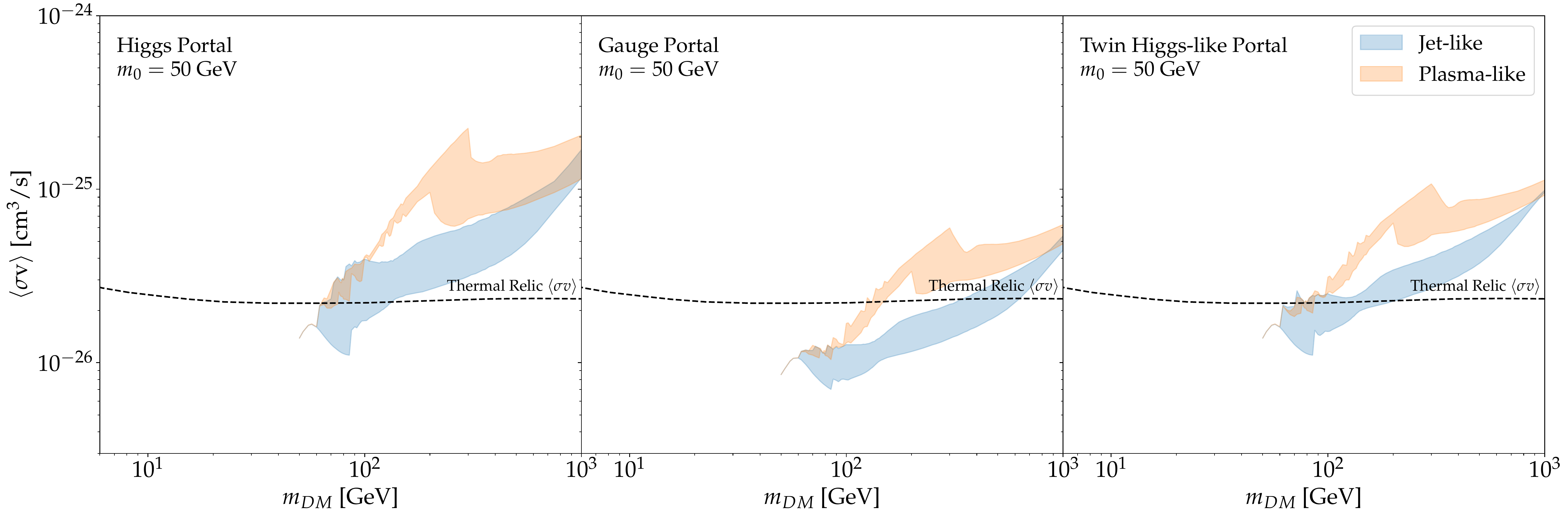}
\end{tabular}
\caption{AMS-02 antiproton limits at 95\% C.L. for DM annihilation to dark glueball showers with $m_0 = 10$ GeV (top) and 50 GeV (bottom). We plot the limits for jet-like (blue) and plasma-like (orange) showers separately for each of the decay portals we consider. The thermal relic cross section is the black dashed line \cite{Steigman:2012nb}.}
\label{fig:constraints_antiproton}
\end{figure}

Figure \ref{fig:constraints_photon_10GeV} shows the 95\% confidence level limits found by comparing our DM photon spectra to dwarf spheroidal galaxy data. Here we focus on $m_0$ values 10 and 50 GeV as they bracket the motivated rage for neutral naturalness theories. For $m_0 = 10$ GeV, across all decay portals we consider, jet-like showers are more constrained compared to the plasma-like case. Thus even with plasma-like showers resulting in higher multiplicity glueball events, the softer spectra leads to weaker constraints. Comparing decay portal cases, the presence of an active dimension-8 operator also leads to stronger constraints. This is to be expected as this operator leads to the production of high energy gamma ray emission, compared to the softer spectra produced in hadronic showers present in dimension-6 operator decays.

Looking at the behaviour of the confidence limits themselves, as expected we also see the jet-like and plasma-like constraints converge to the same limit as $m_{DM}$ approaches $m_0$. This is because in this limit, regardless of hadronization assumption, by energy conservation the shower can only produce two $0^{++}$ glueballs. As $m_{DM}$ begins to increase, initially the shower limits remain similar across benchmarks, as they still can only produce two glueballs, but now producing the heavier states. However above $m_{DM} = 2 m_0$ the jet-like and plasma-like limits begin to diverge as hadronization assumptions start to play a role on spectra evolution.
The jet-like shower limit quickly widens as hadronization dependent behaviour becomes influential. When all jet-like benchmarks have passed their hadronization scale, the jet-like limit maintains a relatively constant width and curve. At large energies the bottom of the band begins to plateau, which is a known feature of dwarf spheroidal constraints \cite{Fermi-LAT:2016uux}.
For the plasma-like shower, the limit quickly weakens compared to the jet-like case. This is because the shower is still in the limit that it is hadronizing as a single plasma that then emits thermal momenta glueballs. Thus even though $m_{DM}$ is increasing, the glueballs are only produced with on average a constant thermal momenta, leading to weakening constraints. However, once $2m_{DM} > \Lambda_{had}$, the gluons perturbatively split before plasma production, leading to boosted plasma and final state glueballs with increasing energy. In this regime you see the limit change in behaviour and approach the jet-like limit. For heavier glueballs with $m_0 = 50$ GeV, the constraints are slightly weakened compared to the light glueball limit for the same $m_{DM}$. Due to this fact and only probing DM masses above 50 GeV, the constraints do not exclude thermal relic DM of any mass.

We also point out that even with considering the eight various hadronization benchmarks, the resultant jet-like and plasma-like constraint bands have a modest thickness and lie mostly on top of each other. 
Therefore, Fermi-LAT Dwarf Spheroidal limits are fairly insensitive to the unknown details of glueball hadronization.

\subsection{AMS-02 Antiprotons}
\label{subsec:antiprotons}

Figure \ref{fig:constraints_antiproton} shows the 95\% confidence limits found by comparing our DM antiproton spectra to AMS-02's antiproton data, again for the motivated bracketing glueball masses $m_0 = 10$ GeV and 50 GeV. We immediately see that these constraints are generally much stronger than the dwarf spheroidal galaxy constraints, which is expected since dark glueballs predominately produce hadronic showers over direct gamma rays. It is also clear that unlike the Dwarf Spheroidal gamma ray limits, antiproton limits are much more sensitive to the unknown aspects of dark glueball hadronization for intermediate glueball masses, and the jet- or plasma-like assumptions lead to qualitatively different behaviour.

While looking quite different than the dwarf spheroidal constraints, many aspects of the previous analysis are still applicable. Some new behaviour is that in the limit $m_{DM}$ approaches $m_0$, the constraints significantly weaken. This is due to the fact the showers produced are so soft they are in the energy regime highly dependent on solar modulation effects. Since we only fit the spectra above 4 GeV to remain conservative about solar modulation assumptions, we lose sensitivity for this $m_{DM}$ range.

Again, until around $m_{DM} = 2 m_0$, both limits remain relatively inline, but past this point begin to diverge drastically as the plasma-like limits severely weaken again. This feature is again due to the antiproton spectra being produced with the same average thermal momenta while the dark matter mass increases. However, once the hadronization scale is saturated, at 40 and 60 GeV respectively, the glueball energy begins to increase and the limits increase in strength, approaching the jet-like constraints in the large $m_{DM}$ limit. The $m_0 = 50$ GeV case follows the same qualitative behaviour.

\subsection{AMS-02 Positrons}
\label{subsec:positrons}

\begin{figure}[t]
\begin{tabular}{c}
\includegraphics[width=\textwidth]{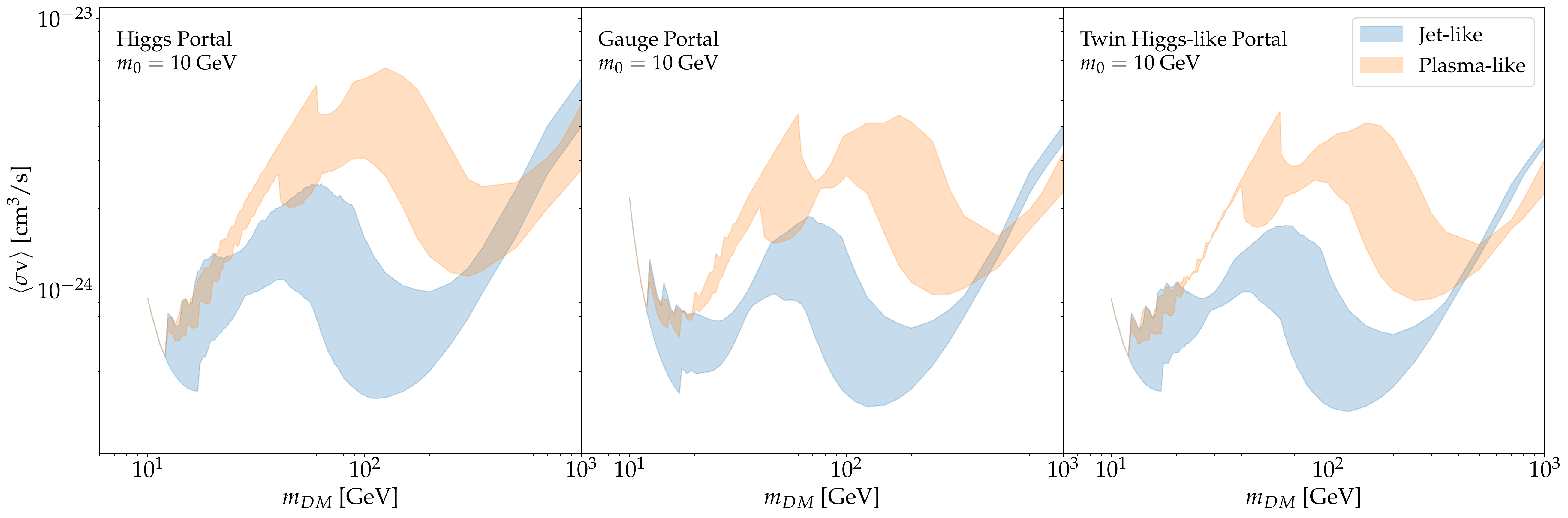}
\\
\includegraphics[width=\textwidth]{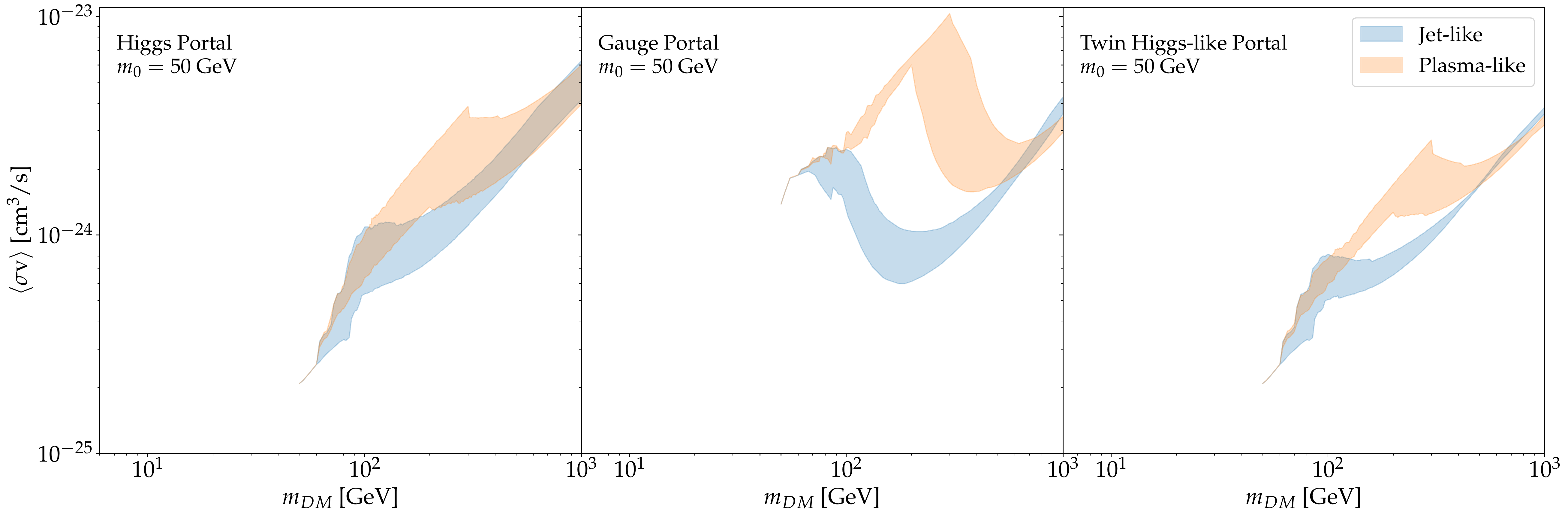}
\end{tabular}
\caption{AMS-02 positron limits at 95\% C.L. for DM annihilation to dark glueball showers with $m_0 = 10$ GeV. We plot the limits for jet-like (blue) and plasma-like (orange) showers separately for each of the decay portals we consider. The thermal relic cross section is the black dashed line \cite{Steigman:2012nb}.}
\label{fig:constraints_positron_10GeV}
\end{figure}

\begin{figure}[t]
\centering
\begin{tabular}{cc}
\begin{turn}{90} \phantom{blablablablabl} \scriptsize Higgs Portal \end{turn}
&
\includegraphics[width=0.84\textwidth]{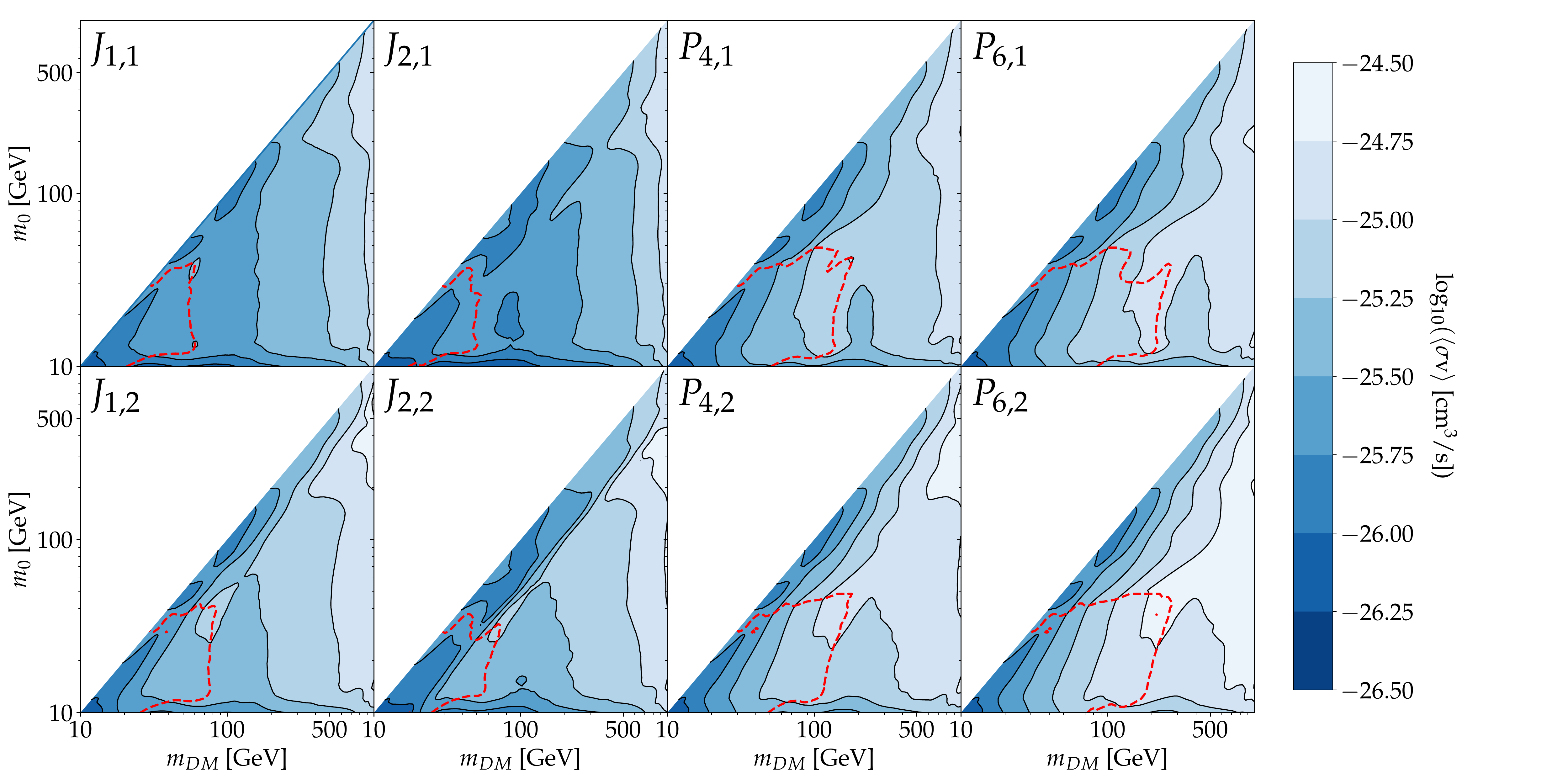}
\\
\begin{turn}{90} \phantom{blablablablabl} \scriptsize  Gauge Portal \end{turn} &
\includegraphics[width=0.84\textwidth]{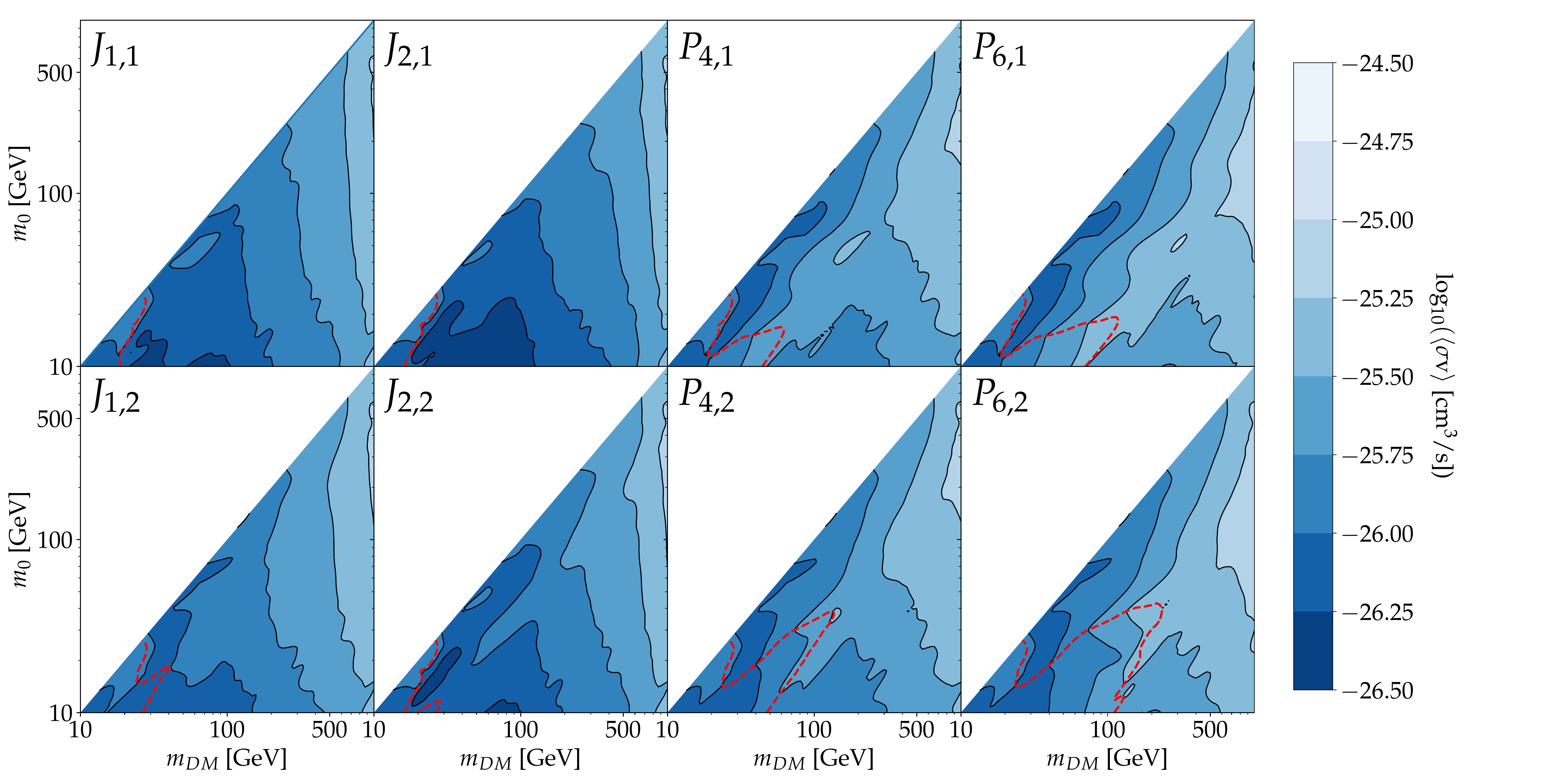}
\\
\begin{turn}{90} \phantom{blablablablabl}  \scriptsize  Twin-Higgs-like \end{turn} &
\includegraphics[width=0.84\textwidth]{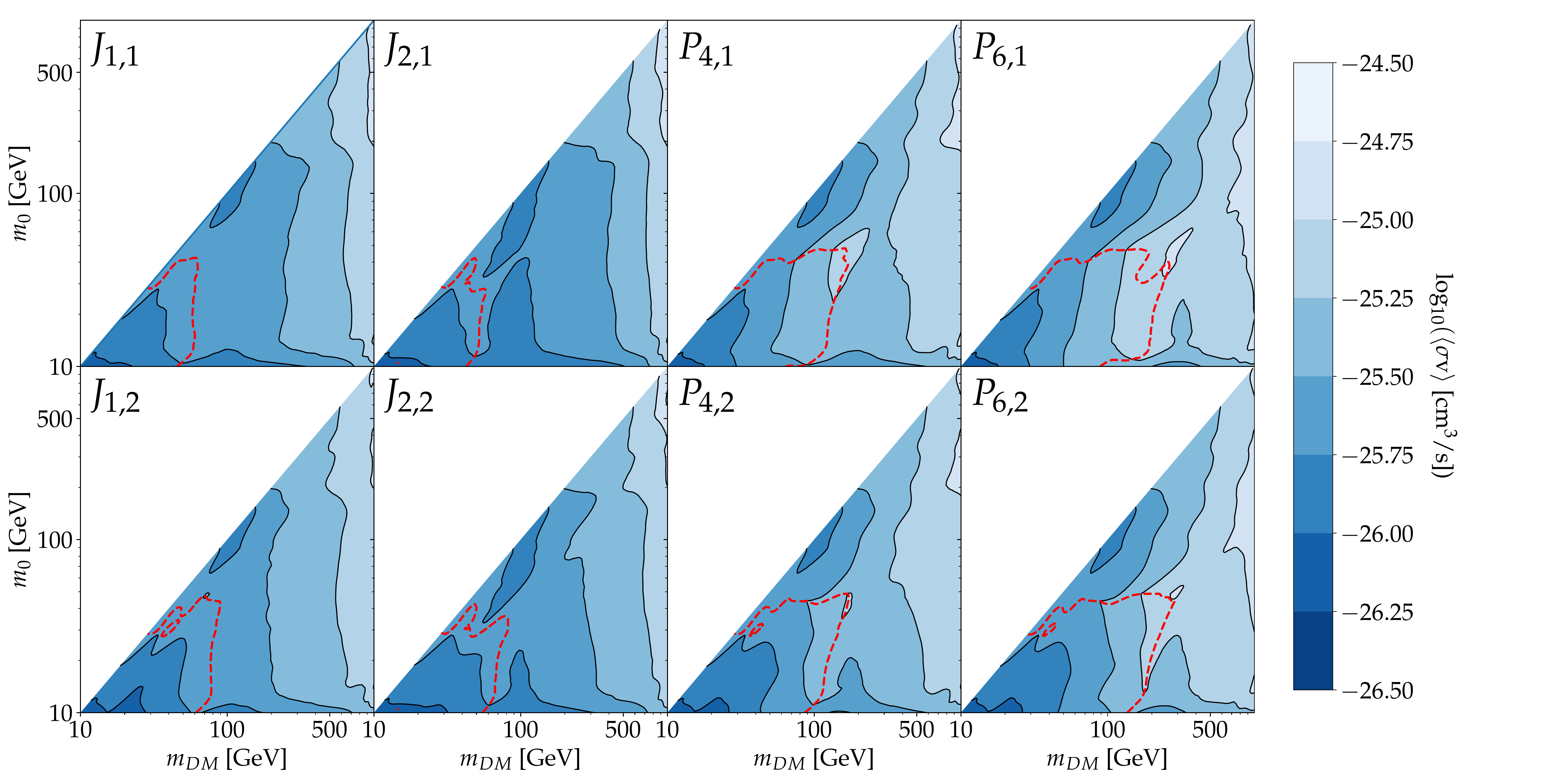}
\end{tabular}
\caption{Strongest indirect detection constraints for Higgs portal (top), gauge portal (middle), and Twin Higgs-like decays (bottom). For each decay portal we show constraints for each of the 8 hadronization benchmarks in Table.~\ref{tab:hadronization_table}. Red dotted line indicates where the strongest constraint switches from  Fermi-LAT (lower left) to AMS-02 antiprotons (upper right).}
\label{fig:superplot}
\end{figure}

Figure \ref{fig:constraints_positron_10GeV}  shows the 95\% confidence level limits found by comparing our DM positron spectra to AMS-02's positron data, for both light and heavy dark glueballs. These limits are included for completeness, as they do not provide any competitive limits compared to both the photon and antiproton channels. A unique feature of the positron limits however is that the plasma-like showers provide stronger limits for larger DM masses. In this regime the plasma-like spectra have a larger energy flux in the low energy bins, compared to the jet-like case, thus providing stronger constraints.

\subsection{Combined Constraints}
\label{subsec:combined}

In this section we summarise the constraints provided by all observation experiments and present them in the full $(m_{DM}, m_0)$ mass plane in Figure \ref{fig:superplot} for all decay portals and hadronization benchmarks.  We plot the strongest constraint for each point in our parameter space, which generally is the AMS-02 antiproton limit, except for the small $m_0$ and $m_{DM}$ region. In this region the strongest constraint is provided by the Fermi-LAT dwarf galaxy limits. These constraints should be applicable to a wide range of complex dark sectors.

In general, increasing the DM mass results in weakened constraints. This is a common feature for indirect detection limits as increasing the DM mass decreases the number density and thus the annihilation rate. We also find that jet-like hadronization leads to stronger constraints. This shows that even with the increased final multiplicity of plasma-like showers, the softer energy spectra weakens bounds.

In most cases, for a constant DM mass, the constraints are fairly insensitive to increases in the  glueball mass scale. This holds true until the lightest glueball mass approaches the dark matter mass. In this limit the final glueball multiplicity drastically decreases, leading to harder and more energetic final states, leading to stronger constraints. In the absolute limit that $m_{DM} = m_0$ the only hadronization possibility is the shower producing two $0^{++}$ glueballs, thus independent of hadronization parameters. We see that across the eight hadronization benchmarks we consider, the constraints agree in this limit for each portal.

\begin{figure}[t]
\begin{tabular}{c}
\includegraphics[width=\textwidth]{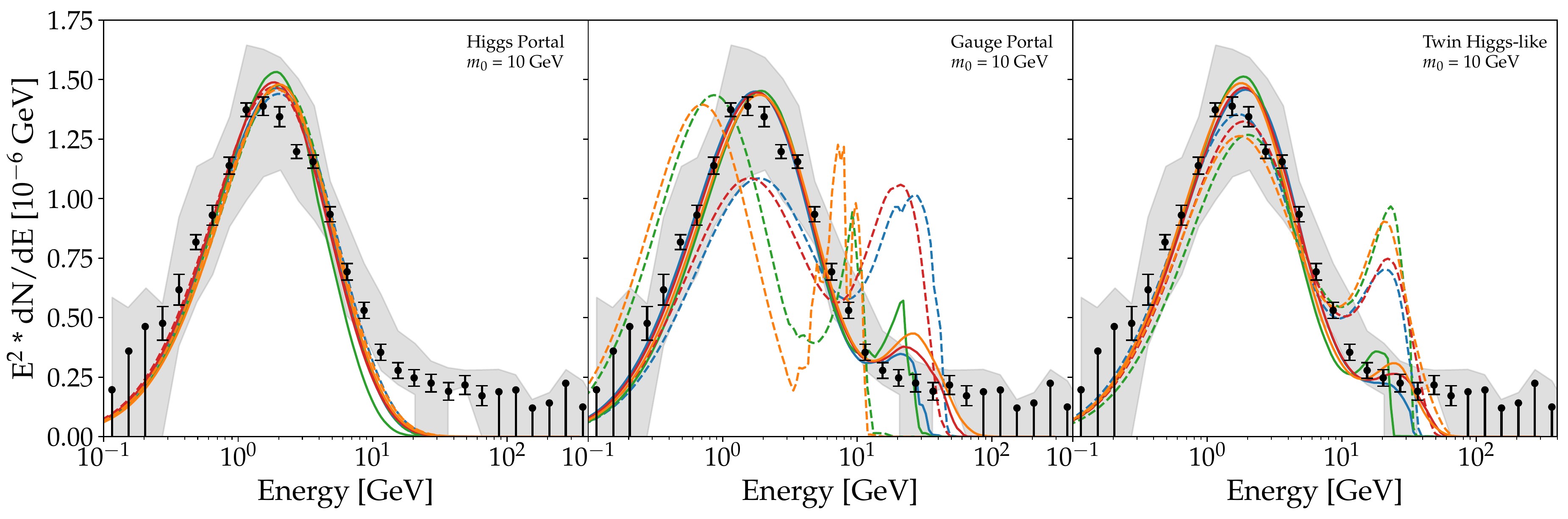}
\\
\includegraphics[width=\textwidth]{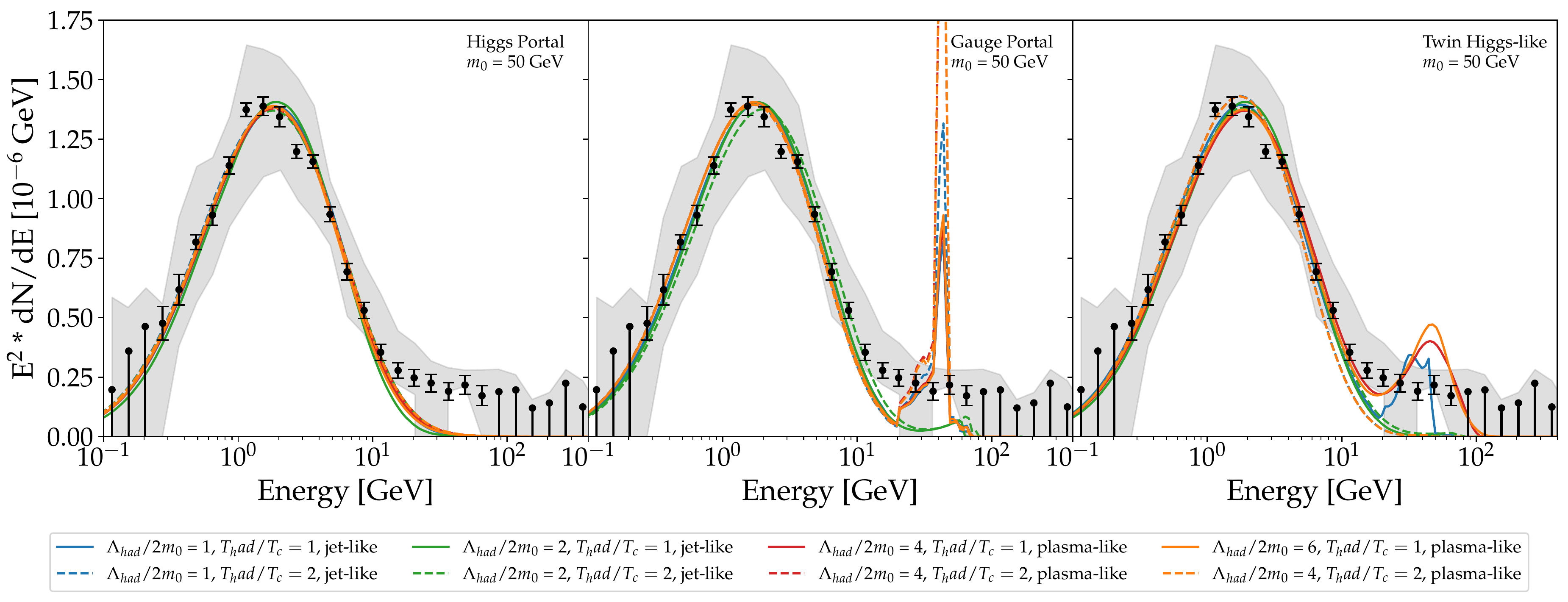}
\end{tabular}
\caption{Best fit photon spectra to the GCE excess, with $m_0 = 10$ GeV (top) and 50 GeV (bottom). The GCE spectra is calculated in Ref. \cite{DiMauro:2021raz}, with statistical error bars or upper limits in black. The grey envelope represents the variation from using different analysis techniques and interstellar emission models.}
\label{fig:GCE_10GeV}
\end{figure}

Comparing the decay portals, we see the strongest constraints are provided by the gauge portal, especially for small DM masses. The Higgs portal and Twin-Higgs like decay portals are quite similar, as expected due to their similar decay channels. However, the Twin-Higgs like case has slightly stronger constraints due to the increased SM flux from the additional $0^{-+}$ and $1^{+-}$ decays.

\section{Constraining a detected dark sector via multimessenger analysis}
\label{sec:multichannel}

\begin{table}
\centering
\resizebox{\textwidth}{!}{
\begin{tabular}{|c|c|c|c|c|}
\hline
UV operator & Hadronization & $m_{DM}$ [GeV] & $\langle\sigma v\rangle$ [$\times 10^{-26}$ cm$^3$ s$^{-1}$] & $\chi^2$/d.o.f. \\
& Benchmark & & & \\ \hline
Higgs Portal & $J_{1,1}$  & 51.5 (83.3) & 1.32 (2.64) & 11.82 (5.79) \\ \cline{2-5} 
 & $J_{1,2}$ & 67.0 (114.7) & 2.32 (5.41) & 9.69 (5.02) \\ \cline{2-5} 
 & $J_{2,1}$ & 38.3 (88.0) & 0.97 (2.38) & 17.08 (4.73) \\ \cline{2-5} 
 & $J_{2,2}$ & 48.8 (98.11) & 1.88 (4.63) & 13.38 (3.45) \\ \cline{2-5} 
 & $P_{4,1}$ & 105.5 (83.8) & 2.83 (2.66) & 10.59 (5.34) \\ \cline{2-5} 
 & $P_{4,2}$ & 94.6 (104.0) & 3.48 (4.94) & 10.50 (6.55) \\ \cline{2-5} 
 & $P_{6,1}$ & 162.8 (84.1) & 4.44 (2.66) & 10.50 (5.62) \\ \cline{2-5} 
 & $P_{6,2}$ & 143.6 (103.9) & 5.92 (4.85) & 9.93 (5.65) \\ \hline
Gauge Portal & $J_{1,1}$ & 46.8 (79.3) & 1.12 (2.03) & 5.05 (3.45) \\ \cline{2-5} 
 & $J_{1,2}$ & 63.7 (79.4) & 1.30 (2.05) & 68.21 (4.15) \\ \cline{2-5} 
 & $J_{2,1}$ & 36.0 (72.2) & 0.83 (1.80) & 9.26 (6.27) \\ \cline{2-5} 
 & $J_{2,2}$ & 52.7 (91.1) & 0.89 (2.30) & 122.93 (4.53) \\ \cline{2-5} 
 & $P_{4,1}$ & 90.9 (79.3) & 2.20 (2.04) & 4.13 (3.91) \\ \cline{2-5} 
 & $P_{4,2}$ & 82.3 (78.7) & 1.66 (2.05) & 71.17 (4.33) \\ \cline{2-5} 
 & $P_{6,1}$ & 150.4 (79.3) & 3.60 (2.04) & 5.47 (3.89) \\ \cline{2-5} 
 & $P_{6,2}$ & 116.7 (78.7) & 2.00 (2.05) & 123.61 (4.17) \\ \hline
Twin Higgs-like & $J_{1,1}$ & 48.3 (79.8) & 1.07 (2.13) & 5.98 (3.82) \\ \cline{2-5} 
 & $J_{1,2}$ & 61.4 (59.1) & 1.34 (1.53) & 16.70 (5.89) \\ \cline{2-5} 
 & $J_{2,1}$ & 36.2 (88.0) & 0.78 (2.34) & 9.79 (4.24) \\ \cline{2-5} 
 & $J_{2,2}$ & 47.4 (87.9) & 0.96 (2.36) & 37.15 (2.91) \\ \cline{2-5} 
 & $P_{4,1}$ & 97.6 (105.8) & 2.19 (2.85) & 4.42 (4.88) \\ \cline{2-5} 
 & $P_{4,2}$ & 82.4 (59.1) & 1.78 (1.53) & 20.54 (6.51) \\ \cline{2-5} 
 & $P_{6,1}$ & 157.9 (80.1) & 3.51 (2.14) & 4.66 (4.52) \\ \cline{2-5} 
 & $P_{6,2}$ & 119.3 (59.1) & 2.50 (1.53) & 33.47 (6.47) \\ \hline
\end{tabular}
}
\caption{Best fit values for DM parameters found by fitting to the GCE spectra reported in Ref. \cite{DiMauro:2021raz}. These values correspond to lightest glueball mass, $m_0$ = 10 (50) GeV.}
\label{tab:bestfit10}
\end{table}

In the previous section we calculated constraints for the DM cross section from Fermi Dwarf-Spheroidal and antiproton observations. Now we also assume that the GCE is a signal from DM annihilating to dark glueballs. Under this assumption, we can further constrain the DM parameter space to regions that provide a good fit to the signal. By combining this fit with the previously calculated constraints, we can examine if certain possibilities for the dark sector properties are excluded by consistency. In all following figures we plot the $3\sigma$ contour for the GCE best fit.

Fig. \ref{fig:GCE_10GeV} depicts the best fit spectra across the various hadronization benchmarks and decay portals. Table. \ref{tab:bestfit10} lists the chi-squared values per degree of freedom for each fit with lightest glueball mass $m_0$ = 10 GeV and 50 GeV. 
We find that Higgs portal spectra are able to provide a relatively good fit. For $m_0 = 10$~GeV, all best fit $\chi^2$/d.o.f values across the hadronization benchmarks lie in the range 9.93 - 17.08. The quality of fit improves for $m_0$ = 50 GeV, with best fit values residing in the range 3.45 - 6.55. For each hadronization benchmark, the simulated spectra are able to fit the main peak of the GCE, but lack the high energy tail that is also present.

In comparison, the gauge and Twin Higgs-like portals are able to provide a better fit to the high energy tail, but only for the low hadronization temperature ($T_{had} = T_c$) showers. For $m_0$ = 10 GeV, these low temperature $\chi^2$/d.o.f values lie in the range 4.13 - 9.26 and 4.42 - 9.79 respectively. The high temperature showers are unable to provide a good fit due to the presence of a large second spectral peak, causing increased $\chi^2$/d.o.f values in the range 68 - 123 and 16.70 - 37.15. For this reason we will discount them from the rest of our analysis. The $m_0 = 50$ GeV case is similar in the fact that the high temperature showers exhibit a second, sharp peak that does not fit the shape of the high energy tail. 
 
\begin{figure}[t]
\hspace*{-1cm}
\begin{tabular}{cc}
\begin{turn}{90} 
\phantom{blablabla} Higgs Portal, $m_0 = 10$~GeV
\end{turn}
&
\includegraphics[width=\textwidth]{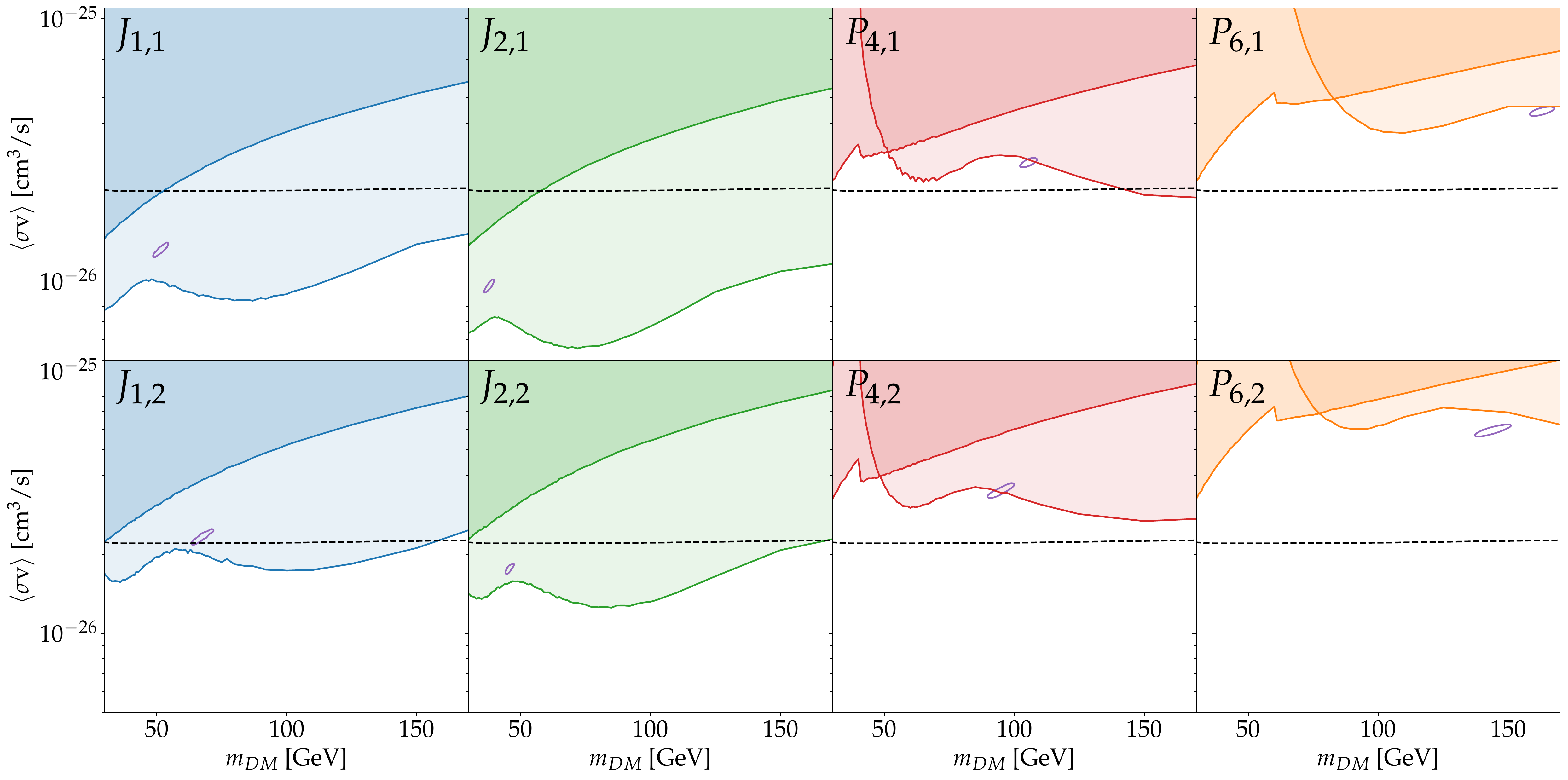}
\\
\\
\begin{turn}{90} 
\phantom{blablabla} Higgs Portal, $m_0 = 50$~GeV
\end{turn}
&
\includegraphics[width=\textwidth]{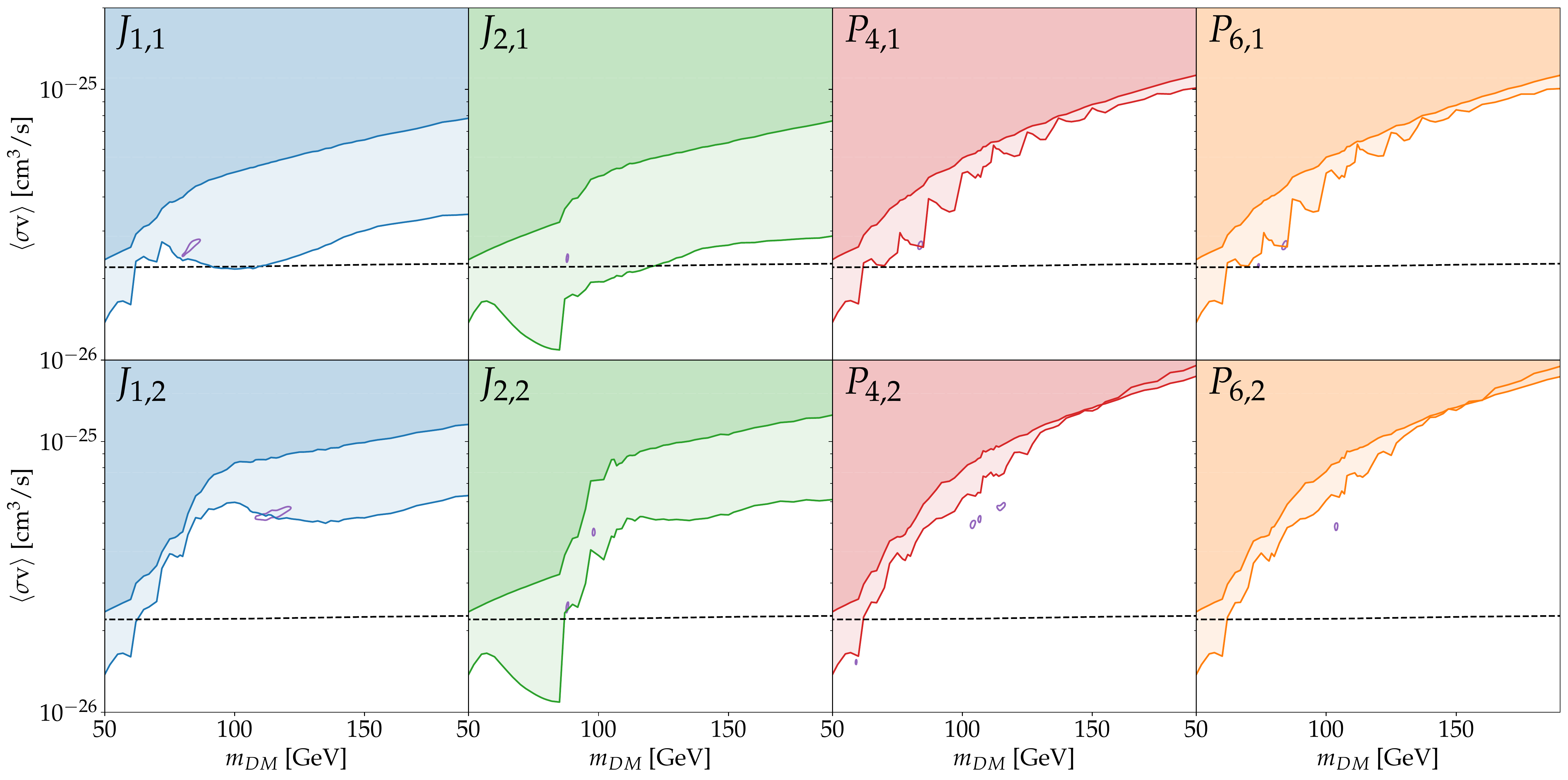}
\end{tabular}
\caption{Combined plot for the Higgs portal decays, with $m_0 = 10$ GeV (top) and 50 GeV (bottom), containing limits at 95\% C.L from both AMS-02 antiproton data (solid) and Fermi-LAT dwarf spheroidal data (hatched). The $3\sigma$ contour for the best fit to the GCE is also depicted in purple for each hadronization benchmark.  The thermal relic cross section is the black dashed line~\cite{Steigman:2012nb}.}
\label{fig:combined_10GeV_dim6}
\end{figure}

In Figure \ref{fig:combined_10GeV_dim6} we plot the favoured DM parameter space in the $m_{DM} - \langle\sigma v\rangle$ plane and the previously calculated constraints for the Higgs portal case. For the jet-like showers, the antiproton constraints exclude the GCE parameter space. Thus we can exclude these hadronization benchmarks as they cannot consistently explain both the GCE and AMS-02 antiproton measurements. Alternatively the plasma-like showers manage to have parameter space that fits the GCE and evades AMS-02 antiproton constraints.

\begin{figure}[t]
\hspace*{-1cm}
\begin{tabular}{cc}
\begin{turn}{90} 
\phantom{blablabla} Gauge Portal, $m_0 = 10$~GeV
\end{turn}
&
\includegraphics[width=\textwidth]{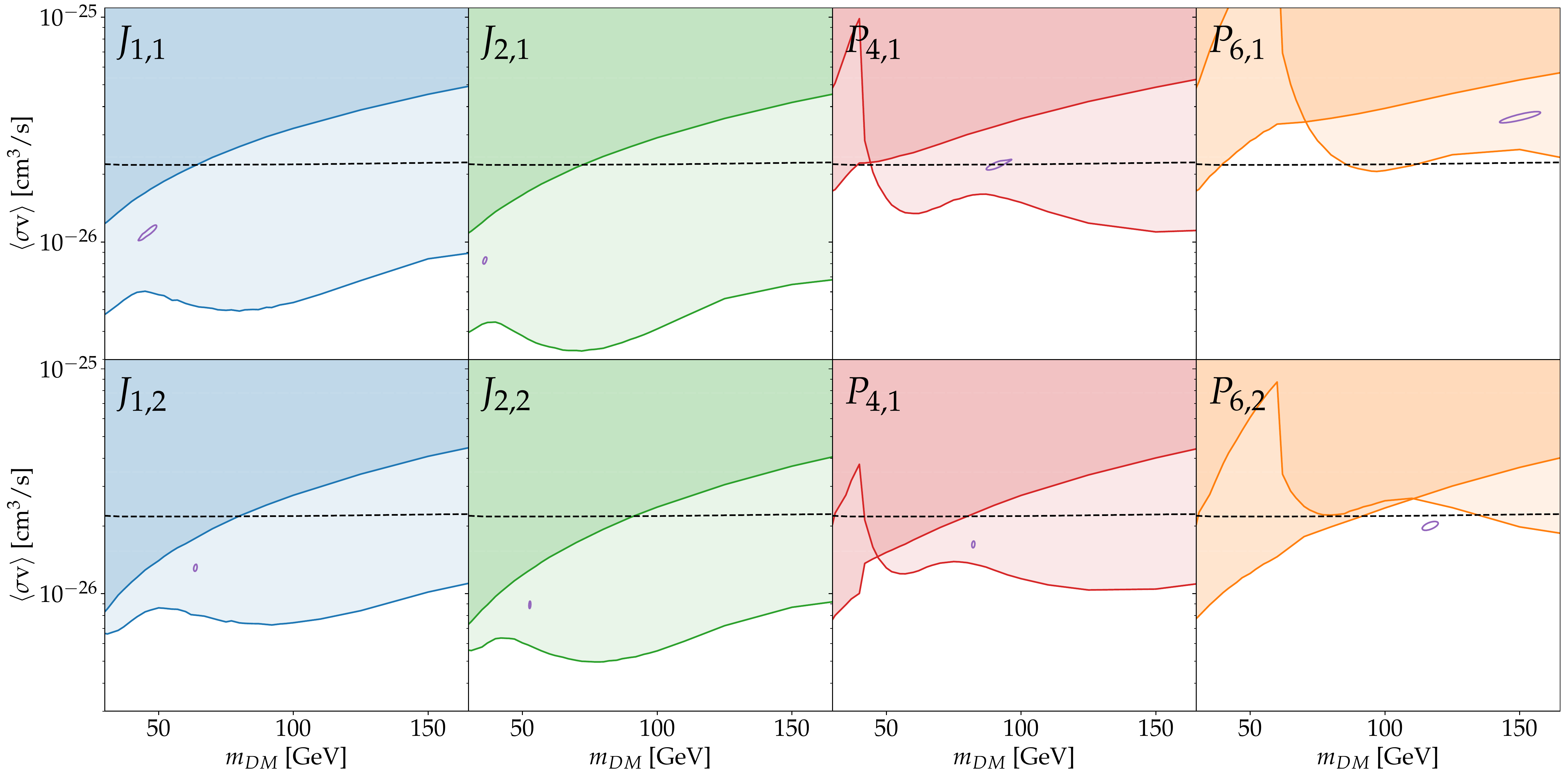}
\\
\\
\begin{turn}{90} 
\phantom{blablabla} Gauge Portal, $m_0 = 50$~GeV
\end{turn}
&
\includegraphics[width=\textwidth]{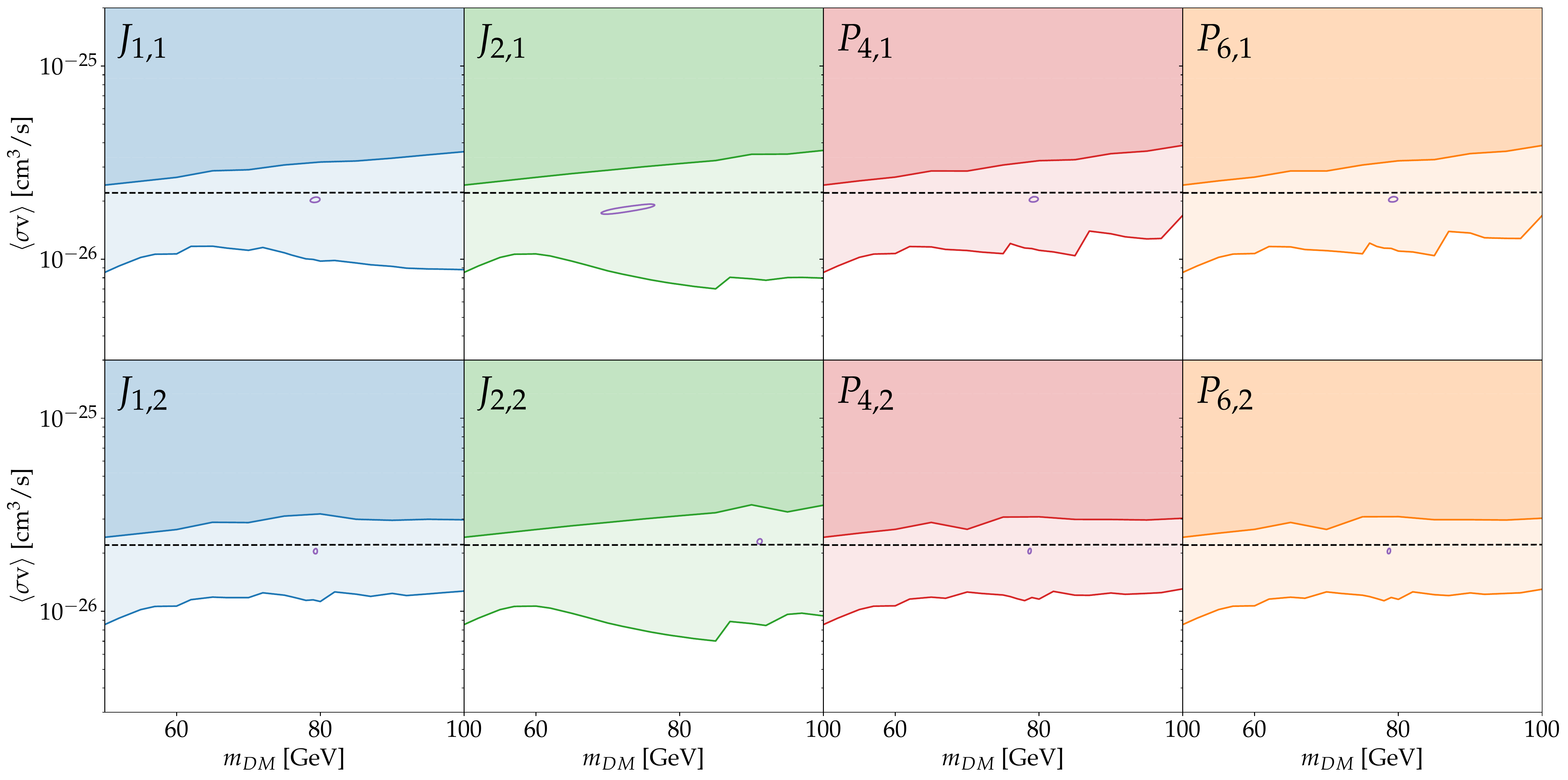}
\end{tabular}
\caption{Combined plot for the gauge portal decays, with $m_0 = 10$ GeV (top) and 50 GeV (bottom), containing limits at 95\% C.L from both AMS-02 antiproton data (solid) and Fermi-LAT dwarf spheroidal data (hatched). The $3\sigma$ contour for the best fit to the GCE is also depicted in purple for each hadronization benchmark. The thermal relic cross section is the black dashed line~\cite{Steigman:2012nb}.}
\label{fig:combined_10GeV_dim8}
\end{figure}

In Figure \ref{fig:combined_10GeV_dim8} we now plot the same favoured parameter space and constraints for the gauge portal case. Considering only the low temperature showers that provide a good fit to the GCE, all parameter space is excluded by the antiproton limits. The same is found for $m_0 = 50$ GeV. Thus for all hadronization benchmarks we consider, the gauge portal case is entirely excluded for these glueball masses. At higher glueball masses the position of the energy spectra peak shifts to higher values and no longer provides a good fit to the GCE. Therefore we reach the same conclusion for these larger $m_0$ values.

\begin{figure}[t]
\hspace*{-1cm}
\begin{tabular}{cc}
\begin{turn}{90} 
\phantom{blablabla} Twin-Higgs-like, $m_0 = 10$~GeV
\end{turn}
&
\includegraphics[width=\textwidth]{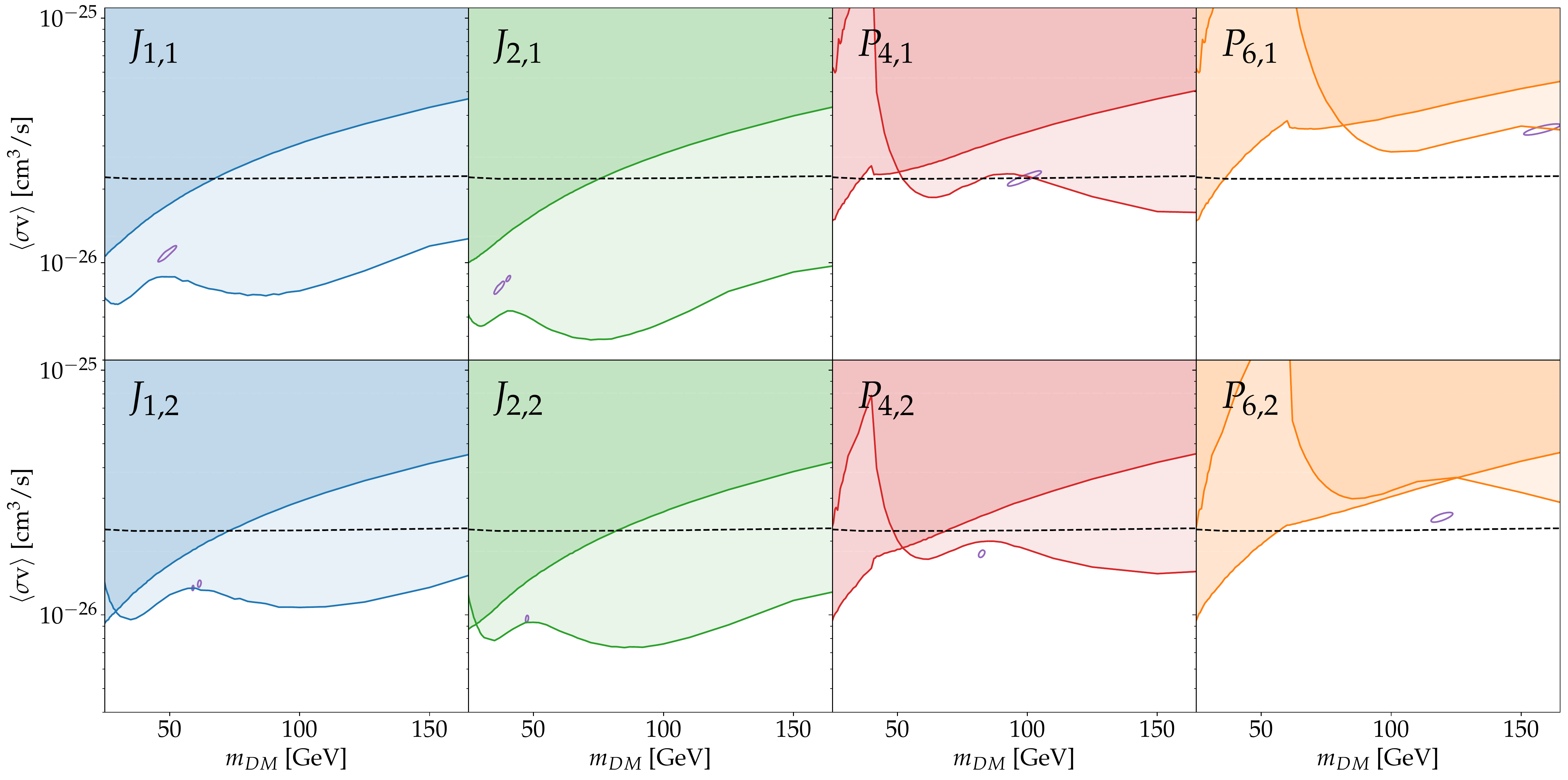}
\\ \\ 
\begin{turn}{90} 
\phantom{blablabla} Twin-Higgs-like, $m_0 = 50$~GeV
\end{turn}
&
\includegraphics[width=\textwidth]{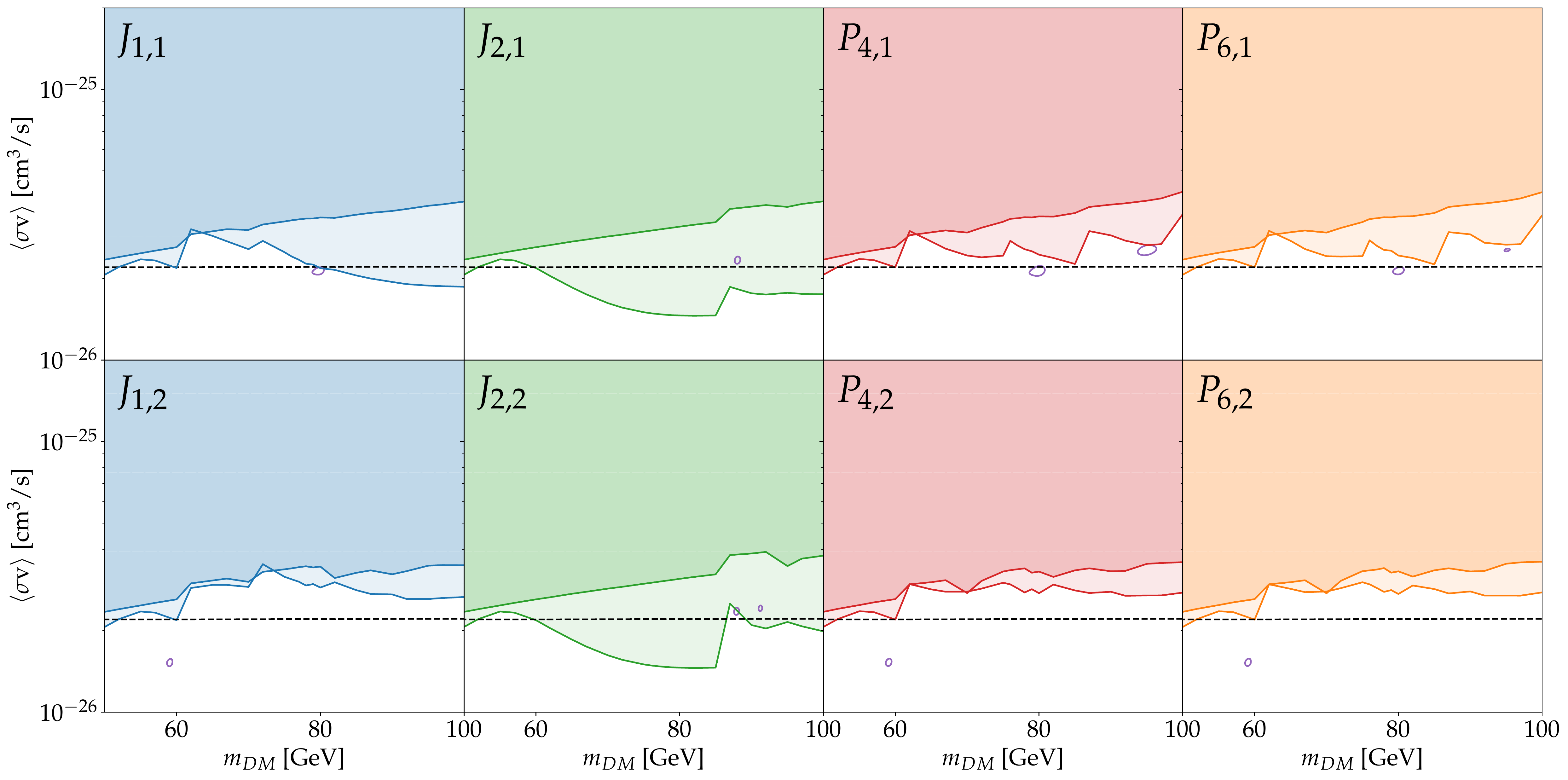}
\end{tabular}
\caption{Combined plot for the Twin Higgs-like decays, with $m_0 = 10$ GeV (top) and 50 GeV (bottom), containing limits at 95\% C.L from both AMS-02 antiproton data (solid) and Fermi-LAT dwarf spheroidal data (hatched). The $3\sigma$ contour for the best fit to the GCE is also depicted in purple for each hadronization benchmark. The thermal relic cross section is the black dashed line~\cite{Steigman:2012nb}.}
\label{fig:combined_10GeV_dim6and8}
\end{figure}

Lastly, in Figure \ref{fig:combined_10GeV_dim6and8} we show the equivalent plot for the Twin Higgs-like case. Again, only the low temperature showers provide a good fit to the GCE and are further considered. For $m_0 = 10$ GeV, both jet-like and plasma-like showers are still allowed, with even the GCE favoured parameter space for $P_{4,1}$ agreeing with the thermal relic cross section. For $m_0 = 50$ GeV, the jet-like shower begins to be constrained as $J_{2,1}$ no longer can the GCE and the antiproton limits. This trend continues for larger glueball masses with the favoured parameter space continuing to be constrained. We check glueball masses for values above the motivated range, 10 - 50 GeV, and find that all low temperature showers get excluded as the spectral peak eventually shifts to larger energies and can no longer reasonably fit the GCE. 

It is interesting to note that dark matter annihilating into glueballs decaying to SM particles via the Twin-Higgs-like combination of Higgs and Gauge portals can explain the GCE with a near-thermal annihilation cross section while being consistent with other constraints.
Antiproton data in particular seems to favour glueball masses in the lower range near 10 GeV, which is motivated by RG arguments for Fraternal Twin Higgs models~\cite{Curtin:2015fna}.
This supports the compelling possibility of a WIMP-like dark matter candidate embedded in the Twin Higgs framework, for example a $\sim 50 - 100$~GeV mirror tau~\cite{Curtin:2021spx}.

\section{Conclusions}
\label{sec:conclusions}

Complex dark sectors are extremely motivated models that are able to provide dark matter candidates as well as mechanisms to solve the Little Hierarchy problem. A general consequence of these models is the existence of long lived particles that are being targeted by LHC searches  \cite{ATLAS:2013bsk,CMS:2018bvr,Alimena:2019zri,ATLAS:2019tkk,CMS:2021dzg} and proposed dedicated external LLP detectors \cite{MATHUSLA:2018bqv, MATHUSLA:2020uve, Curtin:2018mvb, Aielli:2019ivi, FASER:2018bac, Bauer:2019vqk}.
In this work we show that cosmic ray searches for dark matter annihilating into dark glueballs can provide unique and complementary handles for both discovery and diagnosis of such dark sectors. 
In the regime where some or all of these long lived particles decay outside of our detectors, information about their decay is lost.  Since indirect detection probes astrophysical length scales, there is a greater chance to see the entire spectrum of decays, especially if the dark sector contains a range of particles of varying lifetimes, as is the case for dark glueballs. 

We utilised recent progress on simulating pure glue hadronization~\cite{Curtin:2022tou} to present the first indirect detection constraints on dark matter annihilating to dark glueballs that decay to the Standard Model, see Figure~\ref{fig:superplot}. We provide these constraints taking into account uncertainties arising from the unknown nature of pure-glue hadronization, and for three choices of portals via which the dark glueballs decay: Higgs portal, gauge portal and a Twin-Higgs-like mixture, corresponding to different physics in the UV. 
Despite the uncertainties, we derive robust constraints on the masses of dark matter and dark glueballs, with photon and antiproton constraints probing thermal annihilation cross sections. 

Perhaps our most interesting finding is that multi-messenger observation of dark matter annihilation can be used to effectively constrain and even diagnose the properties of the dark sector beyond the masses of the dark matter and dark glueball.
We found that the DM photon spectra are most sensitive to the dark glueball decay portals, while the DM antiproton spectra are most sensitive to hadronization behaviour. 
Figures~\ref{fig:combined_10GeV_dim6}, \ref{fig:combined_10GeV_dim8} and \ref{fig:combined_10GeV_dim6and8} show best-fit regions of the GCE alongside Dwarf Spheroid and antiproton constraints for the Higgs portal, gauge portal and Twin-Higgs-like case respectively. 
Gauge portal decays seem to be entirely excluded from explaining the GCE. On the other hand, the Higgs portal and in particular the Twin-Higgs-like portal are consistent with generating the GCE for thermal annihilation cross sections and  plasma-like dark hadronization behaviour, especially for $\mathcal{O}(10~\mathrm{GeV})$ glueball masses that are favoured in Fraternal Twin Higgs models.

The possibility of a thermal WIMP-like dark matter candidate embedded in a Neutral Naturalness framework is compelling for many reasons~\cite{Craig:2015xla, GarciaGarcia:2015fol, Curtin:2021spx}. Our work shows for the first time that this scenario is explicitly compatible with cosmic ray constraints and can specifically account for the GCE. 
If this scenario were realized by nature, then a plethora of complementary collider and astrophysical observations could allow us to detect dark matter, the dark sector as a whole and its connection to the little Hierarchy problem, an exceedingly exciting prospect. 
This also motivates understanding other subtle signatures of such confining dark sectors in more detail, like the enhanced  production of anti-nuclei~\cite{Winkler:2022zdu}, which could explain the observation of eight antihelium events by AMS-02 \cite{vonDoetinchem:2020vbj}, a smoking gun signal for DM~\cite{Donato:1999gy,Baer:2005tw}.
However, the more general lesson of our study is just as optimistic: complex dark sectors produce complex indirect detection signatures, with enough handles to extract many aspects of the dark sector physics that may be inaccessible at colliders, providing unique and complementary discovery and diagnostic potential. 
With future telescopes such as GAPS \cite{GAPS:2017hga} launching in the near future, indirect detection searches for dark matter are about to enter a very exciting era.

\acknowledgments

The authors especially thank Tim Linden for providing code used in calculating the Fermi-LAT constraints, and Pedro De la Torre Luque for helpful correspondence regarding his customised DRAGON2 code. The authors are also grateful to Daniele Gaggero, Rebecca Leane, David Maurin, Kathrin Nippel, Michael Spira for many insightful conversations.
The research of DC and CG was supported in part by a Discovery Grant from the Natural Sciences and Engineering Research Council of Canada, the Canada Research Chair program, the Alfred P. Sloan Foundation, the Ontario Early Researcher Award, and the University of Toronto McLean Award. The work of CG was also supported by the University of Toronto Connaught International Scholarship, McDonald Institute Graduate Student Exchange program, and the Canada
First Research Excellence Fund.

\bibliography{References}
\bibliographystyle{JHEP}

\end{document}